\PassOptionsToPackage{unicode}{hyperref}
\PassOptionsToPackage{hyphens}{url}
\PassOptionsToPackage{dvipsnames,svgnames,x11names}{xcolor}
\documentclass[
]{article}

\usepackage{amsmath,amssymb}
\usepackage{iftex}
\ifPDFTeX
  \usepackage[T1]{fontenc}
  \usepackage[utf8]{inputenc}
  \usepackage{textcomp} % provide euro and other symbols
\else % if luatex or xetex
  \usepackage{unicode-math}
  \defaultfontfeatures{Scale=MatchLowercase}
  \defaultfontfeatures[\rmfamily]{Ligatures=TeX,Scale=1}
\fi
\usepackage{lmodern}
\ifPDFTeX\else  
\fi
\IfFileExists{upquote.sty}{\usepackage{upquote}}{}
\IfFileExists{microtype.sty}{% use microtype if available
  \usepackage[]{microtype}
  \UseMicrotypeSet[protrusion]{basicmath} % disable protrusion for tt fonts
}{}
\makeatletter
\@ifundefined{KOMAClassName}{% if non-KOMA class
  \IfFileExists{parskip.sty}{%
    \usepackage{parskip}
  }{% else
    \setlength{\parindent}{0pt}
    \setlength{\parskip}{6pt plus 2pt minus 1pt}}
}{% if KOMA class
  \KOMAoptions{parskip=half}}
\makeatother
\usepackage{xcolor}
\ifx\paragraph\undefined\else
  \let\oldparagraph\paragraph
  \renewcommand{\paragraph}[1]{\oldparagraph{#1}\mbox{}}
\fi
\ifx\subparagraph\undefined\else
  \let\oldsubparagraph\subparagraph
  \renewcommand{\subparagraph}[1]{\oldsubparagraph{#1}\mbox{}}
\fi

\usepackage{longtable,booktabs,array}
\usepackage{calc} % for calculating minipage widths
\usepackage{etoolbox}
\makeatletter
\patchcmd\longtable{\par}{\if@noskipsec\mbox{}\fi\par}{}{}
\makeatother
\IfFileExists{footnotehyper.sty}{\usepackage{footnotehyper}}{\usepackage{footnote}}
\makesavenoteenv{longtable}
\usepackage{graphicx}
\makeatletter
\def\maxwidth{\ifdim\Gin@nat@width>\linewidth\linewidth\else\Gin@nat@width\fi}
\def\maxheight{\ifdim\Gin@nat@height>\textheight\textheight\else\Gin@nat@height\fi}
\makeatother
\setkeys{Gin}{width=\maxwidth,height=\maxheight,keepaspectratio}
\makeatletter
\def\fps@figure{htbp}
\makeatother
\NewDocumentCommand\citeproctext{}{}

\makeatletter
 \let\@cite@ofmt\@firstofone
 \def\@biblabel#1{}
 \def\@cite#1#2{{#1\if@tempswa , #2\fi}}
\makeatother
\newlength{\cslhangindent}
\newlength{\csllabelwidth}
\newenvironment{CSLReferences}[2] % #1 hanging-indent, #2 entry-spacing
 {\begin{list}{}{%
  \setlength{\itemindent}{0pt}
  \setlength{\leftmargin}{0pt}
  \setlength{\parsep}{0pt}
  \ifodd #1
   \setlength{\leftmargin}{\cslhangindent}
   \setlength{\itemindent}{-1\cslhangindent}
  \fi
  \setlength{\itemsep}{#2\baselineskip}}}
 {\end{list}}
\usepackage{calc}

\usepackage{booktabs}
\usepackage{longtable}
\usepackage{array}
\usepackage{multirow}
\usepackage{wrapfig}
\usepackage{float}
\usepackage{colortbl}
\usepackage{pdflscape}
\usepackage{tabu}
\usepackage{threeparttable}
\usepackage{threeparttablex}
\usepackage[normalem]{ulem}
\usepackage{makecell}
\usepackage{xcolor}
\usepackage{orcidlink}
\usepackage{amsmath}
\usepackage[T1]{fontenc}
\usepackage[ruled,vlined]{algorithm2e}
\usepackage{setspace}
\usepackage{hyperref}
\usepackage{longtable}
\usepackage{float}
\usepackage{pdflscape}
\usepackage[capitalise,noabbrev]{cleveref}
\usepackage{caption}
\usepackage{subcaption}
\usepackage{xcolor}
\usepackage{csvsimple}
\usepackage{mathtools}
\makeatletter
\@ifpackageloaded{caption}{}{\usepackage{caption}}
\AtBeginDocument{%
\ifdefined\contentsname
  \renewcommand*\contentsname{Table of contents}
\else
  \newcommand\contentsname{Table of contents}
\fi
\ifdefined\listfigurename
  \renewcommand*\listfigurename{List of Figures}
\else
  \newcommand\listfigurename{List of Figures}
\fi
\ifdefined\listtablename
  \renewcommand*\listtablename{List of Tables}
\else
  \newcommand\listtablename{List of Tables}
\fi
\ifdefined\figurename
  \renewcommand*\figurename{Figure}
\else
  \newcommand\figurename{Figure}
\fi
\ifdefined\tablename
  \renewcommand*\tablename{Table}
\else
  \newcommand\tablename{Table}
\fi
}
\@ifpackageloaded{float}{}{\usepackage{float}}
\floatstyle{ruled}
\@ifundefined{c@chapter}{\newfloat{codelisting}{h}{lop}}{\newfloat{codelisting}{h}{lop}[chapter]}
\floatname{codelisting}{Listing}

\makeatother
\makeatletter
\@ifpackageloaded{caption}{}{\usepackage{caption}}
\@ifpackageloaded{subcaption}{}{\usepackage{subcaption}}
\makeatother
\ifLuaTeX
  \usepackage{selnolig}  % disable illegal ligatures
\fi
\usepackage{bookmark}

\IfFileExists{xurl.sty}{\usepackage{xurl}}{} % add URL line breaks if available
\hypersetup{
  pdftitle={A Random Forest approach to detect and identify Unlawful Insider Trading},
  pdfauthor={Krishna Neupane; Igor Griva},
  pdfkeywords={unlawful insider trading, ensemble methods, decision
trees, random forest, fradulent behavior},
  colorlinks=true,
  linkcolor={blue},
  filecolor={Maroon},
  citecolor={Blue},
  urlcolor={Blue},
  pdfcreator={LaTeX via pandoc}}

\usepackage{setspace}
\title{A Random Forest approach to detect and identify Unlawful Insider
Trading}
\def\asep{\\\\\\ } % default: all authors on same column
\author{\textbf{Krishna
Neupane}~\orcidlink{0000-0003-3911-3988}\\Department of Computational
and Data Sciences\\George Mason University\\Fairfax,
VA\\\href{mailto:kneupan@gmu.edu}{kneupan@gmu.edu}\asep\textbf{Igor
Griva}\\Department of Mathematical Sciences\\George Mason
University\\\\}
\date{}
\begin{document}
\maketitle
\begin{abstract}
According to \emph{The Exchange Act}, \(1934\) unlawful insider trading
is the abuse of access to privileged corporate information. While a
blurred line between ``routine'' the ``opportunistic'' insider trading
exists, detection of strategies that insiders mold to maneuver fair
market prices to their advantage is an uphill battle for hand-engineered
approaches. In the context of detailed high-dimensional financial and
trade data that are structurally built by multiple covariates, in this
study, we explore, implement and provide detailed comparison to the
existing study (Deng et al. (2021)) and independently implement
automated end-to-end state-of-art methods by integrating principal
component analysis to the random forest (\emph{PCA-RF}) followed by a
standalone random forest (\emph{RF}) with \(320\) and \(3984\) randomly
selected, semi-manually labeled and normalized transactions from
multiple industry. The settings successfully uncover latent structures
and detect unlawful insider trading. Among the multiple scenarios, our
best-performing model accurately classified \(96.43\) percent of
transactions. Among all transactions the models find \(95.47\) lawful as
lawful and \(98.00\) unlawful as unlawful percent. Besides, the model
makes very few mistakes in classifying lawful as unlawful by missing
only \(2.00\) percent. In addition to the classification task, model
generated Gini Impurity based features ranking, our analysis show
ownership and governance related features based on permutation values
play important roles. In summary, a simple yet powerful automated
end-to-end method relieves labor-intensive activities to redirect
resources to enhance rule-making and tracking the uncaptured unlawful
insider trading transactions. We emphasize that developed financial and
trading features are capable of uncovering fraudulent behaviors.
\end{abstract}
{\bfseries \emph Keywords}
\def\sep{\textbullet\ }
unlawful insider trading \sep ensemble methods \sep decision
trees \sep random forest \sep 
fradulent behavior

\section{Introduction}\label{sec-introduction}

Corporate insiders, \emph{ex officio}, have privileged access to
material non-public preferential information (MNPI). Primed by
informational advantage, access to and exploitation of \emph{MNPI}
prowess insiders to maneuver fair market price deviate markedly (Grubbs
(1969)) to profit (Binz and Graham (2022), Ahern (2018), Chemmanur, He,
and Hu (2009), Easley, Hvidkjaer, and O'Hara (2002), Amihud (2002), Kyle
(1985)). Behooved as ``routine'' (Cohen, Malloy, and Pomorski (2012)),
firmed on MNPI anomalous transactions are motley strategies of momentum,
value and growth, investment, profitability, intangibles and trading
frictions (K. Hou, Xue, and Zhang (2020)). Unlike heavily
hand-engineered approaches deprived by flexibility and poor replications
(Mayo and Hand (2022)), the machine learning methods distill from
underlying complex data distribution to identify features to discover
unexpected patterns (Varol et al. (2017)). Various statistical
techniques are proposed to detect anomalous behaviors (Kahneman,
Knetsch, and Thaler (1991)) each relevant to the inquiry. In machine
learning, prominent methods are based on proximity (e.g., k-means,
expectation maximization) and distance (e.g., principal component
analysis, k-means) (Breunig et al. (2000), Salgado et al. (2016)). For
example, Rizvi, Attew, and Farid (2022) implements Kernel-based
Principal Component Analysis to analyze the cause and effect of extreme
price movements; Seth and Chaudhary (2020), Goldberg et al. (2003) and
Rizvi, Belatreche, and Bouridane (2019) separately use supervised
Natural language Processing (e.g.~multi-stage SVM) to classify unlawful
insider transactions as a result of the news cycle and other events;
Islam, Khaled Ghafoor, and Eberle (2018) and Li et al. (2022) integrate
decision trees with deep learning (LSTM RNN) to classify transactions
based on the public release of non-public information on the news. A
synthesis of current research in unlawful insider trading is:

\begin{itemize}
\item
  Econometric studies (Cerniglia and Fabozzi (2020), John and Narayanan
  (1997), Fishman and Hagerty (1995)) aim to capture time-varying
  properties insider trading confining to ``trading size'' is an example
  data snooping (also known as specification searches). Therefore
  leaving aside a multitude of other interactive features (e.g.~official
  designations, information arrival timing) to limit into empirical
  irregularities and inconsistencies (Leamer (1978)).
\item
  Statistical methods (Auto Regressive Moving Average (ARIMA) (H. M.
  Anderson (2007), Lutkepohl (1993), McKenzie (1984))), Box and Jenkins
  (Box, Jenkins, and MacGregor (1972)), Hidden Markov Models (Rabiner
  and Juang (1986)) and general class of linear models (Hand (2009),
  Hamilton (1989)) \emph{lack scalability} with growing number of
  observations (\(n\)). These methods are evaluated in a single set of
  train/test splits and are prone to over-generalization (Ge and Smyth
  (2000)).
\item
  Legal scholars lack consensus (Chang et al. (2022), Bainbridge (2022),
  Ma and Sun (1998), Machan (1996),Macey (1988), Manne (1966)) reject
  ``unfairness in market'' favoring full legal protection of insider
  activities to spur innovation while other oppose scholars (Ali and
  Hirshleifer (2017), Gangopadhyay and Yook (2022), Cumming, Johan, and
  Li (2011)) claiming insider trading impedes investor confidence,
  increases agency costs and exacerbates opportunistic managerial
  behaviors.
\end{itemize}

More recently, machine learning techniques have been introduced to
identify and characterize unlawful insider trading (see for example Deng
et al. (2021) and Deng et al. (2019) (collectively hereinafter referred
as \emph{DCZ}), and Lauar and Arbex Valle (2020)). These studies
implement ensemble methods, namely, Random Forest (\emph{RF}) and
extreme Gradient Boosting (XGBoost) to detect unlawful insider trading
showcasing improvement of the classification accuracy. In fraud
detection (money laundering, credit card fraud) these tools improve the
robustness of normal behavior modeling accomplished by parameter weight
updates and recalibration for the majority voting from the ensemble of
weak learners to enhance detection accuracy (Sundarkumar and Ravi
(2015), Louzada and Ara (2012)). However, these methods, are not
resistant to limitations, illustratively, bias-variance trade-off when
exposed to multiple categorical (Strobl et al. (2007)) or continuous
variables (Altmann et al. (2010)). A better control of the model
complexity can be achieved by increasing the tree depth though that
results in increased variance and reduced bias. There are several
advantages of ensemble methods. First, efficient ensemble techniques
fuse data mining and modeling to extract predictive features in a
unified framework in cheaper, faster and creative ways of \emph{data
labeling} to automate routine tasks (Iskhakov, Rust, and Schjerning
(2020)) by uniquely positioning to capture \emph{short-run noise}
features that contribute to historical context and derive meaningful
inferences leading to \emph{trustworthy} predictive power\footnote{Prediction
  is defined as the ability to take known information to generate new
  information(Agrawal, Gans, and Goldfarb (2019))} for
\emph{understandable inferences} (Khademian (2022), Igami (2018),
Christensen, Hail, and Leuz (2016)). Second, domain agnostic methods
such as decision trees and neural networks explain the
\emph{micro-data-structure} of data independent estimators (Malhotra
(2021), Iskhakov, Rust, and Schjerning (2020)) and automatically
generate relatively important variables.

The study contributes first by addressing gaps of hand-engineered,
omitted variables, inter-dependency and multi-dimensionality problems of
data to understand, explain and replicate the anomalous transactions (K.
Hou, Xue, and Zhang (2020)) by training ensemble based models to
discover regularities that uncover the choices that insiders make
(Camerer (2019), Fudenberg and Liang (2019)). Second, identify patterns
that may have been missed by \emph{p-hacking}\footnote{``A focus on
  novel, confirmatory, and statistically significant results leads to
  substantial bias in the scientific literature. One type of bias, known
  as ``p-hacking,'' occurs when researchers collect or select data or
  statistical analyses until nonsignificant results become
  significant.'' (Head et al. (2015))} and detect and characterize
hidden patterns of insider trading strategies including fraudulent
trades based on the performance of financial proxies.

We organize our study in the following sections as follows:
Section~\ref{sec-method-proposed-method} introduces the proposed
methodology outlining the theory behind the principal component analysis
(Section~\ref{sec-method-principal-component-analysis}) and random
forest (Section~\ref{sec-method-random-forest}), the tuning parameters
(Section~\ref{sec-method-parameter-turning}), feature selection criteria
(Section~\ref{sec-method-feature-importance}) and performance measures
(Section~\ref{sec-method-performance-measure});
Section~\ref{sec-analysis-experimental-setup} describes the experimental
setup such as data-preprocessing steps, implementation of
cross-validation, criteria of parameter control and integration of
methos; Section~\ref{sec-analysis-data-results} analyzes the results
that begin with data description
(Section~\ref{sec-analysis-data-results}), steps of dimensionality
reduction (Section~\ref{sec-analysis-dimensionality-reduction}), results
from components of confusion matrix
(Section~\ref{sec-analysis-Results-Classification-Transactions}),
ranking of important and relevant variables
(Section~\ref{sec-analysis-variable-importance}), and ranking of
variables (Section~\ref{sec-analysis-variable-importance-impurity-based}
and
Section~\ref{sec-analysis-variable-importance-permutation-importance}).
In the final two sections- we discuss results
(Section~\ref{sec-analysis-discussions}) and provide our conclusions,
recommendations and future course
(Section~\ref{sec-conclusions-future}).

\section{Proposed Methodology}\label{sec-method-proposed-method}

Multicovariate and high dimensional financial and trade data are
structurally built with underlying hidden attributes. To detect
irregularities in data under consideration, methods such as principal
component analysis (Aıt-Sahalia and Xiu (2019)), gradient boosting
(Eggensperger et. al. (2018)), random forest (Svetnik et al. (2003)) and
neural networks (Hilal, Gadsden, and Yawney (2022)) have been proven
effective (Bakumenko and Elragal (2022), Gu, Kelly, and Xiu (2018)).
Notably, these methods act as intermediary steps of the empirical work
in economics (Athey (2019)). Unsupervised method, \emph{PCA} effectively
controls the variability and reduces dimensions
(Section~\ref{sec-method-principal-component-analysis}) and supervised
method, \emph{RF} reduces bias of \emph{decorrelated trees}
(Section~\ref{sec-method-random-forest}). As a first-of-its-kind
application in identifying unlawful insider trading \emph{DCZ}
integrated two methods illustrating to be the best of both worlds. In
analogous settings of \emph{DCZ}, we extend powerful modeling techniques
to reduce data dimensions control covariates and classify unlawful
insider trading exhibiting comparable data characteristics in the US
security market and hence the results are easily compared and assessed.
Subsequently, we extend our experiment to evaluate performance by
implementing \emph{RF} for their success in wider applicability.
Therefore, by resorting to proven methods from previous research and
extending the number of features in the study from \(26\) in \emph{DCZ}
study to \(110\) features, we experiment with \emph{RF} to reduce result
uncertainty, quality decline, bias, unfairness failing systems to
provide transparency, interpretability, and reproducibility of results
(Stoyanovich et al. (2017), O'Neil (2016)). We follow number of
data-preprocessing steps before implementing techniques as explained in
Section~\ref{sec-analysis-experimental-setup}.

\subsection{Principal Component
Analysis}\label{sec-method-principal-component-analysis}

We implement \emph{PCA} to compare our results to \emph{DCZ}. \emph{PCA}
filters noise to decorrelate features uncovering latent dynamics to
extract the most relevant information from the compressed data by
computing principal components- the linear combinations of the original
variables (Wang et al. (2018), Abdi and Williams (2010)). The first
component contains the largest variance and other components are
constrained as orthogonal the first to explain the variance. The values
of the components are geometric projections of observations onto
principal components called factor scores. Among the variety of
successful applications, the domain agnostic \emph{PCA} is applied in
areas of our interest of finance and anomaly detection (Aıt-Sahalia and
Xiu (2019)) to identify exceptional volatilities (Egloff, Leippold, and
Wu (2010)), acknowledge peculiar investor sentiments (M. Baker and
Wurgler (2006)), discern policy uncertainties (S. R. Baker, Bloom, and
Davis (2016)), stock and bond return aberrant (Pasini (2017), Pérignon,
Smith, and Villa (2007), Driessen, Melenberg, and Nijman (2003), Feeney
(1967)) and more recently market cross-correlation and systemic risk
measurement (Billio et al. (2012), Zheng et al. (2012), Kritzman et al.
(2011)). \emph{DCZ} implements \emph{PCA} with \(26\) features and
accomplishes the task of identifying unlawful insider trading in reduced
dimensions. Similarly, we implement \emph{PCA} and obtain comparative
results. Besides, we extend the experiment with additional \(110\)
features. Formally, \(\mathbf{D} \in \mathbf{R} ^{n \times m}\) is
matrix with \(n\) rows and \(m\) features. Given the
\(p-\text{dimensional}\) vectors, performing PCA means summarizing them
by projecting down to \(q\)-dimensional subspace, principal components
sequentially organized in real coordinate space. The first component,
the \(i\)-th vector in the sequence is the direction that is orthogonal
to the second component \(i\)-1 with maximum variance thus minimizing
the average squared perpendicular distance from the points to the line
that in turn reduces the random variability because of computed reduced
dimensions and distances (Aluja-Banet and Nonell-Torrent (1991)). The
steps of the \emph{PCA} algorithm are shown in Algorithm
\ref{alg:principalComponentAnalsyis}. The core component of \emph{PCA}
analysis is an estimation of the eigenvalues and covariance matrix. The
eigenvalues of covariance matrix consistent asymptotic normal estimators
(T. W. Anderson (1963)).

\singlespacing

\begin{algorithm}[H]
  \SetKwInOut{Input}{Input}
  \SetKwInOut{Output}{Output}
  \Input{$\mathbf{D}^{n \times m} $ Data matrix}
  1. Obtain the empirical mean of each column\\
  2. Center the data by subtracting off the mean of each column of $p \times 1$-dimensions, represented as $\mathbf{B}$\\
  3. Compute the covariance matrix $C=\frac{1}{N}\mathbf{B}^T \mathbf{B}$ from above Step 2\\
  4. Compute the eigenvalues and eigenvectors of $\mathcal{C}$ so $\mathbf{V}^{-1}$ $\mathbf{CV} = \mathbf{E}, \mathbf{V}$ holds eigenvectors of $\mathbf{C}$ and $ \mathbf{E} $ is the diagonal $M \times M$ diagonal eigenvalue matrix \\
  5. Sort the columns of $\mathbf{D}$ into order of decreasing eigenvalues\\
  6. Obtain the cumulative values of eigenvalues by summing up eigenvalues for each row (represents the overall predictive power)\\
  7. Obtain the projections with eigenvalues 
 \caption{Principal Component Analysis}
 \label{alg:principalComponentAnalsyis}
\end{algorithm}

\doublespacing

Despite its wide success and acceptability, the method has drawbacks
(Aıt-Sahalia and Xiu (2019)). First, the classical problem of the curse
of dimensionality, the growth of data dimensions(volume) corresponding
to the sample size the algorithms suffer the curse of dimensional
(Bellman (1958)), which typically means the required data needs to grow
exponentially with its dimensions contributing to shallow the
consistency of eigenvalues. Second, parameter constancy, the principal
components are the linear combinations derived from data that
potentially may fail to capture the non-linear data patterns. Third,
shortcomings related to non-stationarity, principal components are based
on data(vector) independence. However, time series data abound of
dependency and lack of stationarity thus inferences become inappropriate
(Zhang and Tong (2022), Brillinger (2001)). By removing noise and
de-correlating data, the contribution and relevancy of prominent
individual features become conspicuous. In identifying unlawful
transactions integrating PCA and \emph{RF} re-calibrates leading
feature's weights thereby minimizing the risk of misclassification and
falsely unlawful activity.

\subsection{Random Forest}\label{sec-method-random-forest}

\emph{DCZ} integrates \emph{PCA}
(Section~\ref{sec-method-principal-component-analysis}) to Random Forest
(\emph{RF}), a data-driven non-parametric ensemble method that
implements bootstrap resampling (bagging)was pioneered by Breiman
(2001). It, therefore, offers the advantage of Out-of-bag (\emph{OOB})
error prediction. The bagging is a sub-sampling method with a
replacement that creates training examples for a model to learn from
before the construction of the next individual tree. The \emph{OOB}
prediction error estimates of the \emph{RF} is the mean prediction error
on each tree sample used if and only if these trees did not participate
in the bootstrap sampling of the base learner (Hastie, Tibshirani, and
Friedman (2009)). In addition, \emph{RF} provides flexibility by
incorporating non-linearities and interactions of the multitude of
decision trees that eventually aggregate de-correlated trees to control
variance and increase accuracy (Chen et al. (2022)). In simple terms,
\emph{RF} is designed to group observations with similar predictors by
growing trees in a sequence of ``steps''. Recent theoretical results
suggest that the \emph{RF} is consistent with the central limit theorem
(Ramosaj and Pauly (2023), Wager and Athey (2018), Svetnik et al.
(2003), Liaw, Wiener, et al. (2002), Breiman (2001)). The method has
been widely adopted in chemical informatics (Svetnik et al. (2003)),
ecology (Prasad, Iverson, and Liaw (2006)), \(3\)-D Object recognition
(Shotton et al. (2011)), epidemiology (Azar et al. (2014)), and remote
sensing (Pal (2005)) to name few.

\(\textbf{Definition}\)

\begin{quote}
A random forest is a classifier consisting of a collection of
tree-structured classifier \(h(x, \Theta_k ), k=1, .....,k\) where the
\({\Theta_k}\) are independently distributed random vectors and each
tree casts a unit vote for the most popular class at input \(x\)
(Breiman (2001)). The random forest algorithm is shown in Algorithm
\ref{alg:randomforest}\footnote{https://pages.cs.wisc.edu/\textasciitilde matthewb/pages/notes/pdf/ensembles/RandomForests.pdf}.
\end{quote}

\begin{algorithm}
  \SetKwInOut{Input}{Input}
  \SetKwInOut{Output}{Output}
  \Input{$\mathbf{D}:=(x_1, y_1),....,(x_n, y_n)$ with Features $\mathbf{F}$, and number of trees in forest $\mathbf{B} $ }
  \Begin{
    \text{RANDOMFOREST}($\mathbf{S}, \mathbf{F}$)\\
    $H \leftarrow \emptyset$\\
    \For {$ i \in 1, ...., \mathbf{B}$}{
      $\mathbf{S}^{i} \leftarrow \text{A bootstrap sample from } \mathbf{S}$\\
      $\mathbf{h}_i \leftarrow$ \text{RANDOMIZEDTREELEARN}
      $(\mathbf{S}^{(i)}, \mathbf{F})$\\
      $H \leftarrow H  \cup H \{h_i\} $
      }
      return H
      }
  \Begin{
    \text{RANDOMIZEDTREELEARN}($\mathbf{S}, \mathbf{F}$)\\
    At each node:\\
     \Indp{$\mathbf{f} \leftarrow$ very small subset of $\mathbf{F}$}\\
     Split on best feature in $f$\\
    \Indm{return The Learned Tree}
  }
   
 \caption{Random Forest}
 \label{alg:randomforest}
\end{algorithm}

In supervised settings with iterative learning, randomness is injected
into \emph{RF} to correctly map and classify input values (\(x\)) to its
target (\(y\)), expressed as functional relationship \(y= f(x)\) for
unknown joint distribution \(P(x,y)\). The decision tree minimizes the
correlation \(\rho\) among the collection of \(p-\) dimensional
variables. The common elements of growing the ensemble \(k\) tree are
that for the \(k^{th}\) tree, a random vector \(\Theta_k\) is generated
independent of its past \(\Theta_1,....., \Theta_k\) but has the same
distribution which results in a classifier \(h( \mathbf{x}, \Theta_k)\)
with \(\mathbf{x}\) as input vectors with \(N\) number of examples in
the training set. \emph{RF} learns about the functional relationship
between \(x\) and \(y\) by producing a prediction model
\(\hat{f}(X, \theta)\), controlled by the \(k\)-dimensional
hyperparameter configuration \(\theta=(\theta_1,... ,\theta_k)\) from
the search space \(\Theta=\Theta_1 \times .... \times \Theta_k\). Doing
so, it estimates the expected risk of the inducing algorithm, w.r.t
\(\theta\) on new data, also sampled from \(P\). The results obtained in
Algorithm \ref{alg:randomforest} are further tuned to extract the
optimized parameter values with \(\textit{k-fold}\) cross-validation to
minimize empirical risk with repeated resampling to approximate the
generalization error (Afendras and Markatou (2015), Bergstra and Bengio
(2012), Witten et al. (2011), Arlot and Celisse (2010), Shao (1993),
Efron (1986), Efron (1983), Stone (1977), Geisser (1975), Stone (1974)).

\subsection{Parameter Tuning}\label{sec-method-parameter-turning}

Parameter tuning reduces the overall \emph{OOB} prediction error.
Shortlist of \emph{RF} parameters to enhance accuracy, minimize data
overfitting and improvise the relative contribution to correctly predict
response are (B, C, and Munoz (2021), Eggensperger
et, al. (2018), Friedman (2001)):

\begin{itemize}
\item
  \(m_{try}\): Number of predictor features drawn randomly as candidates
  for each tree per node to balance the correlation preserving
  independence between trees. To control the split-variable
  randomization, for classification \(m_{tries}\) is calculated as the
  square root of number of features \(\sqrt{n_{features}}\). In summary,
  the lower values of \(m_{tries}\) tend to stabilize aggregation and
  produce ``near optimal'' outcomes by selecting less correlated and
  different trees (Probst, Bischl, and Boulesteix (2018), Bernard,
  Heutte, and Adam (2009)).
\item
  \(ntrees\): Although technically not a hyperparameter, the number of
  trees(\(ntrees\)) stabilizes trees, by, ``\ldots empirically, the
  optimal number of trees is obtained when the forest error reaches its
  limit as the number of trees grows to infinity\ldots{}'' (Probst,
  Bischl, and Boulesteix (2018), Scornet (2017)). Higher \(ntrees\)
  increases computational time.
\item
  \(max \ depth\): Number of splits each decision tree is allowed to
  make. Originally Breiman (2001) proposed to keep the tree unpruned and
  allow it to grow deep (Belkin et al. (2019)), the recent empirical
  results suggest proper tuning can significantly improve the
  performance but has time cost complexity (Probst, Bischl, and
  Boulesteix (2018), Scornet (2017)).
\item
  \(sample \ rate\) has a default value of 1.0 improves the
  generalization error for better predictive powers of validation and/or
  test results
\end{itemize}

\subsection{Feature Importance}\label{sec-method-feature-importance}

The multi covariates input during the tree construction of the \emph{RF}
model fitting aims to extract, compare and rank relatively significant
ones for better interpretability, explainability and predictive accuracy
of the model in the downstream application (Schölkopf et al. (2001))
that reduces time, space complexity and generalization error (S. Zhou
(2022), Xu et al. (2014), Duchi et al. (2008)). Various methods have
been proposed to identify interactive non-linearities to allow the
incorporation of the known sparsity structure (Qian et al. (2022), Xu et
al. (2014), Guyon et al. (2010), Genuer, Poggi, and Tuleau-Malot (2010),
Strobl et al. (2007)). Orginally, Breiman (2001) implemented \emph{Mean
Decrease of Impurity (MDI)} based on \emph{Gini Scores} which is a
decrease in node impurity, that is, the probability of incorrectly
classifying a randomly chosen transactions in the dataset if it were
labeled randomly according to the class distribution (Nembrini, König,
and Wright (2018)). As the technical by-product, in \emph{RF} splitting
rules maximizes impurity reduction introduced by the node-split.
Therefore, a node with larger decrease in impurity value is ranked
higher than the one with lower value. MDI approach has two limitations,
first, it ranks features during the training so it does not have access
to test data and second, it inherently favors highly cardinal features
therefore dilutes the significance of the correlated features.

Subsequent work by Fisher, Rudin, and Dominici (2019) investigates
drawbacks of the \emph{MDI} methods by applying a \emph{permutation}
based approach to calculate the contribution of each \(m^{th}\)
covariate in the out-of-bag examples keeping all other predictor
variables fixed which is passed on to other corresponding trees
(Nembrini, König, and Wright (2018)). The method lowers bias and
balances MDI's high cardinality (typically of numerical features)
favorability over low cardinality features (binary and/or categorical
defined by a smaller number of possible categories). We compare our
results including the influence of correlation between the co-variates
and the number of correlated variables (Gregorutti, Michel, and
Saint-Pierre (2016)) from each of these methods in subsequent sections
(see Section~\ref{sec-analysis-variable-importance}).

\subsection{Performance Measure}\label{sec-method-performance-measure}

To evaluate the performance of the binary supervised classification
problem a squared matrix known as confusion matrix of size
\(2 \times 2\) schematically represented as in
Table~\ref{tbl-confusionMatrixSchematic} is widely used. The confusion
matrix that represents the actual and predicted classes along rows and
columns respectively (Hastie, Tibshirani, and Friedman (2009)).

\singlespacing

\begin{table}

\caption{\label{tbl-confusionMatrixSchematic}A schematic representation
of the components of confuison matrix for binary classifcation of
insider trading transactions}

\centering{

\centering
\begin{tabular}{lll}
\toprule
\multicolumn{1}{c}{  } & \multicolumn{2}{c}{Predicted Labels (PP+PN) } \\
\cmidrule(l{3pt}r{3pt}){2-3}
Tot. Pop. (P+N) & Lawful(P) & Unlawful(N)\\
\midrule
\cellcolor{gray!10}{Lawful(+)} & \cellcolor{gray!10}{True Lawful} & \cellcolor{gray!10}{False Unlawful}\\
Unlawful(-) & False Lawful & True Unlawful\\
\bottomrule
\end{tabular}

}

\end{table}%

\begin{longtable}{>{\hspace{0pt}}m{0.7\linewidth}>{\hspace{0pt}}m{0.25\linewidth}}
\caption{Definition and formula to compute components of the confusion matrix from the Table \ref{tbl-confusionMatrixSchematic}}\\ 

\textbf{Definition} & \textbf{Equation}\\
\hline
True Lawful (True Positives) are lawful transactions classified as lawful, that is, ($+$ as $+$). True Positive Rate (TPR, Recall or Sensitivity), calculated with Equation \ref{eq:eqTruePositiveEquation}, is the proportion of positive instances that are correctly classified as positive by the model. & 

    \begin{equation}
        \begin{aligned}
            \scalebox{0.8}{$\text{TPR}= \frac{TP}{TP+FN}$} \\
            \scalebox{0.8}{$= \frac{TP}{P}$}
            \scalebox{0.8}{$ =1-FNR$}
        \end{aligned}
        \label{eq:eqTruePositiveEquation}
    \end{equation} 
    \\
\hline
False Lawful (False Positives) are unlawful transactions but the model wrongly classifies as lawful, that is, ($-$ as $+$). False Positive Rate (FPR, False Alarm or Fall Out) calculated with Equation \ref{eq:eqFalsePositiveEquation}, is the probability that incorrect classification of unlawful as lawful. & 
    \begin{equation}
        \begin{aligned}
            \scalebox{0.8}{$\text{FPR}= \frac{FP}{FP+TN}$} \\
            \scalebox{0.8}{$ = \frac{FP}{N}$} 
            \scalebox{0.8}{$ =1-TNR$}
        \end{aligned}
        \label{eq:eqFalsePositiveEquation}
    \end{equation}
 \\ 
\hline
True Unlawful (True Negatives) represent unlawful transactions correctly identified as unlawful, that is, ($-$ as $-$). True Negative Rate (TNR, Specificity) is calculated as in Equation \ref{eq:eqTrueNegativeEquation}. &
    \begin{equation}
        \begin{aligned}
           \scalebox{0.8}{$\text{TNR} = \frac{TN}{TN+FP}$} \\ 
            \scalebox{0.8}{$= \frac{TN}{N}$}
            \scalebox{0.8}{$=1-FPR$}
        \end{aligned}
        \label{eq:eqTrueNegativeEquation}
    \end{equation}
\\ 
\hline
False Unlawful (False Negatives) are lawful transactions, but model wrongly classifies as unlawful(negative), that is, ($+$ as $-$). False Negative Rate (FNR, Miss Rates) calculated with Equation \ref{eq:eqFalseNegativeEquation} is the  probability of true negatives (lawful) missed equation. &
    \begin{equation}
        \begin{aligned}
           \scalebox{0.8}{$\text{FNR}= \frac{FN}{TP+FN}$}\\ 
            \scalebox{0.8}{$= \frac{FN}{P}$}
            \scalebox{0.8}{$ =1-TPR$} 
        \end{aligned}
        \label{eq:eqFalseNegativeEquation}
    \end{equation}
\\
\hline
Accuracy (ACC) measures effectiveness of classifier to the correctly identify transactions across all lawful($+$) and unlawful($-$) transactions. In unbalanced dataset the presence of large negative class and the measure favors of the true negatives, the outcomes become inconsequential Equation (see \ref{eq:eqAccuracyEquation}). &
    \begin{equation}
        \begin{aligned}
           \scalebox{0.8}{$\text{ACC} = \frac{TP + TN}{P + N}$} \\ 
           \scalebox{0.8}{$= \frac{TN+TP}{TP+FP+FN+TN}$}
        \end{aligned}
        \label{eq:eqAccuracyEquation}
    \end{equation}
\\
\hline
Precision (PRE) is proportion of correct to the sum of the correct and incorrect classification that the class agreement of the positive labels classified by the classifier (see Equation \ref{eq:eqPrecisionEquation}). &
    \begin{equation}
        \begin{aligned}
           \scalebox{0.8}{$ \text{PRE} =\frac{TP}{TP+FP}$}
        \end{aligned}
        \label{eq:eqPrecisionEquation}
    \end{equation}
\\
\hline
\end{longtable}                                                                                           

\onehalfspacing

\section{Experimental Setup}\label{sec-analysis-experimental-setup}

The Exchange Act, \(1934\) requires individual insiders to submit the
Statement of Changes in Beneficial Ownership (Form \(4\)) within
forty-eight hours of the transaction. For the study, we downloaded
\(9.6\) million publicly available individual Form \(4\) from Electronic
Data Gathering, Analysis and Retrieval (EDGAR) between \(2003-2022\),
followed by pre-processing with a customary \emph{XML} python text
crawler which is eventually stored as a document in the \emph{MongoDB}
collection. We use a unique Central Index Key (\emph{CIK}) from
\emph{SEC} to query and merge daily trade and financial results
retrieved respectively from the Center for Research in Security Prices
and Compustat-CapitalIQ. An individual complete \emph{MongoDB} document
consists of \(110\) features representing ownership, corporate
governance, profitability, financial performance, risk and returns in
the security market. We generate unlawful labels based on defendant
names extracted from the publicly available court complaints filed by
the \emph{SEC}. A text crawler integrated with Levenshtein distance
utility extracts and matches the defendant insider to the filers of the
Form \(4\). A transaction is labeled negative (unlawful) if a score is
\(85\) or higher during comparison. Our database consists a total of
\(1992\) unlawful transactions. For the completeness of data, if
quarterly financial results are unavailable, we replace missing data
from the subsequent quarter.

We establish a feature-to-feature comparison with \emph{DCZ}. Our
comparison matches or closely resemble \(20\) numerical covariates (see
Table~\ref{tbl-Variables}) of \emph{DCZ}. We are unable to retrieve
features such as CR5, H5, CR5, and Z-indices that we intentionally
replace with \(5\) categorical features, namely, Acquisition and
Disposition, IsDirector, IsOfficer, and IsTenPercentOwner. Our full
feature space consists \(110\) features from which we are able to subset
\(25\) to match \emph{DCZ}. Notably, these features have long history of
scrutiny and successful adoption in economic and financial models such
as Black-Scholes Model (Black and Scholes (1973)), Capital Asset Pricing
Model (Sharpe (1964)), Efficient Market Hypothesis (Fama (1970)),
Arbitrage Pricing Theory (Ross (1978)) and Behavioral Financial
Economics (Campbell and Shiller (1988), Tversky and Kahneman (1974)).
Additionally, several of these features have been implemented to predict
corporate financial distress (Qian et al. (2022), Sun et al. (2020), L.
Zhou, Lu, and Fujita (2015)). The numerical features in the dataset
\(X_i\) (\(i = 1, ...., n\)) are normalized by \emph{z-score}\footnote{\(\frac{(X_i- \mu)}{\sigma}\).
  The numerator improves the interpretation of the principal effects and
  scaled by the standard deviation in the denominator puts all the
  predictors into a common scale by minimizing the expected differences
  in the outcomes(Gelman (2008)).} for faster convergence and that
follows standard normal distribution \(N(0,1)\) expressed as zero mean
(\(\mu\))\footnote{\(\mu= \frac{1}{n} \sum_{i=1}^n x_i\)} and unit
standard deviation (\(\sigma\))\footnote{\(\sigma = \sqrt{\frac{1}{n} \sum_{i=1}^n(x_i-\mu)^2}\)}.
The \(5\) categorical features are one-hot encoded.

We experiment in the balanced dataset with \(0.5 \colon 0.5\) ratio of
lawful to unlawful respectively for \(320\) and \(3984\) transactions.
The smaller subset is randomly selected from the larger pool. Each of
these experiments is further sub-divided by the number of features
followed by \emph{PCA} integration and without. The \(1992\)
transactions are the maximum number of unlawful that we can randomly
choose from. However lawful transactions are randomly sampled from the
pool of the remaining \(9.6\) million. For instance, to construct a set
of \(320\) transactions we randomly choose \(120\) transactions from
\(1992\) unlawful and remaining transactions from other pools. We
initialize our experiments with pre-defined hyperparameters in a
randomized search space with iterating in \(5-\)fold cross-validation.
We pick \(ntrees\), \(m_{try}\), \(max \ depth\), and \(sample \ rate\)
as hyperparameters. We maintain \(0.8:0.2\) training to testing split
ratio and experiment for in eight experiments repeating one hundred
times. We report results in
Table~\ref{tbl-rfComparativeConfusionMatrixKrishnaResults}.

After fitting the model, we extract and rank important features based on
the reduction of the Gini Impurity. It is well known \emph{MDI} favors
high cardinal features and is derived from the statistics of the
training data. To improve the limitations, we introduce permutation
importance as an additional model inspection technique. Such setting
allows us to control the influence of correlation. As an additional
step, because financial and trade data are naturally exhibit high
collinearity, we adopt a hierarchical clustering based on the Spearman
rank-order correlation, picking a threshold, and a single feature from
each cluster to rank features. In addition, we relaxed certain
assumptions of \emph{DCZ} as follows:

\begin{itemize}
  \item{Qualitative features with prefix such as \textit{significant} are notional descriptors. With the possibility of subjective interpretations, it may inadvertently introduce spuriousness. The \textit{DCZ} study adopts the criteria defined by \textit{significace} without providing further explanation of what it quantitatively meant. Thus, as a precaution, in this study, we limit the usage of qualitative definitions to reduce incongruent elucidation thereby extending the empirical justification of results.}
  \item{Confining data selection to the same industry group is loosely a \textit{selection bias}. In contrast to \textit{DCZ} study, for wider applicability, we randomly selected transactions from different industrial groups thereby boosting objectivity. }
  \item{In the dynamic and complex United States Securities Market, it is conceivable that there exists a plethora of transactions that may go unnoticed despite being unlawful. The authorities lack resources to investigate the majority of transactions, thereby it is fair to treat each transaction as potentially unlawful.}
\end{itemize}

Our code accesses various python based objects of scikit package
repeatedly for \(100\) times. We record performance of each of the
components of confusion matrix in a separate spreadsheet and average at
the end. The repetition aims to the control the variability and more
reliable than a single experiment. Our results are benchmarked against
\emph{ANN}, \emph{SVM}, \emph{Adaboost} and \emph{Random Forest} from
Deng et al. (2019) and Deng et al. (2021).

\section{Results}\label{sec-analysis-data-results}

Our results are based on the balanced dataset with \(3984\) (\(1992\)
unlawful) transactions of \(110\) dimensions as explained in
Section~\ref{sec-method-proposed-method}. For illustrative purposes, we
present Table~\ref{tbl-transactionsCount} (b) as a random example subset
of Table~\ref{tbl-transactionsCount} (a) that matches Deng et al.
(2021). Notably, in each of these tables, the existence of the higher
number of ``purchases'' is because of the executive's compensation
structures, specifically, the restricted stock options and bonuses
allocated towards meeting targets.

\singlespacing

\begin{table}

\caption{\label{tbl-transactionsCount}Illustrative distribution of the
count of labeled Insider Trading Transactions. A maximum of number of
\(1992\) labeled unlawful transactions are available. A balanced dataset
(0.5:0.5, lawful to unlawful ratio) is constructed with additional 1992
lawful transactions. To illustrate 3 (a) displays additional \(1992\)
lawful randomly selected transactions from a pool of 9.5 million. In the
right-hand side (3(b)), a subset of randomly selected \(320\) from (a)
are shown that matches the counts of the benchmark studies - Deng et
al.~(2021) and Deng et al.~(2019).}

\begin{minipage}{0.50\linewidth}

\begin{longtable*}[l]{lrrr}
\toprule
Label & Sell & Purchases & Total\\
\midrule
\cellcolor{gray!10}{Lawful} & \cellcolor{gray!10}{405} & \cellcolor{gray!10}{1587} & \cellcolor{gray!10}{1992}\\
Unlawful & 318 & 1674 & 1992\\
\bottomrule
\end{longtable*}

\end{minipage}%
\begin{minipage}{0.50\linewidth}

\begin{longtable*}[l]{lrrr}
\toprule
Label & Sell & Purchases & Total\\
\midrule
\cellcolor{gray!10}{Lawful} & \cellcolor{gray!10}{27} & \cellcolor{gray!10}{133} & \cellcolor{gray!10}{160}\\
Unlawful & 26 & 134 & 160\\
\bottomrule
\end{longtable*}

\end{minipage}%

\end{table}%

\onehalfspacing

\subsection{Results of Dimensionality
Reduction}\label{sec-analysis-dimensionality-reduction}

In this section, we report results from the principal component analysis
(\emph{PCA}) that transforms the multi-dimensional correlated variables
into a smaller set of the principal components (\emph{PCs}, directions)
computed from the linear combinations capturing the largest variation in
the original dataset. \emph{PC}s are eigenvectors that represent the
weight of eigenvalues of the covariance matrix with the largest
explaining the direction of the most variability explained by the given
component. The first \emph{PC} captures the most variability. The
eigenvector times the square root of the eigenvalue gives the component
loadings which can be interpreted as the correlation of each item with
the \emph{PC}. In order to decide on the number of components required
to explain the proportion of the total variance explained by each
\emph{PC}, we calculate the cumulative explained variance ratio - the
ratio of its eigenvalue to the sum of all the eigenvalues of all
\emph{PC}s measuring the relative contribution of each \emph{PC} to
explain the amount of variance. In our context, we aggregate an average
of \(94.76\) percent cumulative \emph{EVR} explained by \(10\)
\emph{PCs}.

\begin{figure}

\begin{minipage}{0.50\linewidth}

\centering{

\includegraphics{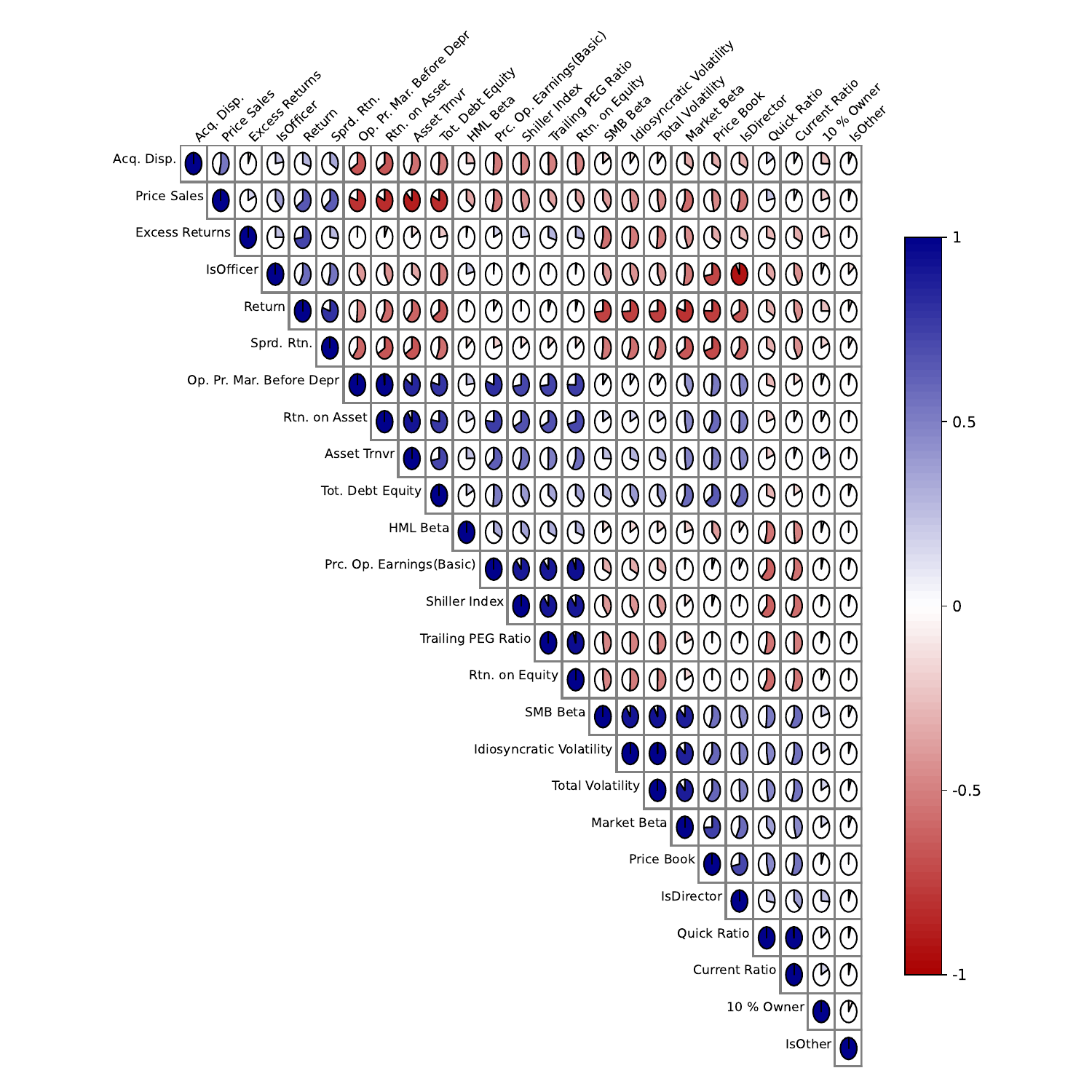}

}

\subcaption{\label{fig-corrleationPlot-withoutPCA}Correlation between
raw features before the model fit}

\end{minipage}%
\begin{minipage}{0.50\linewidth}

\centering{

\includegraphics{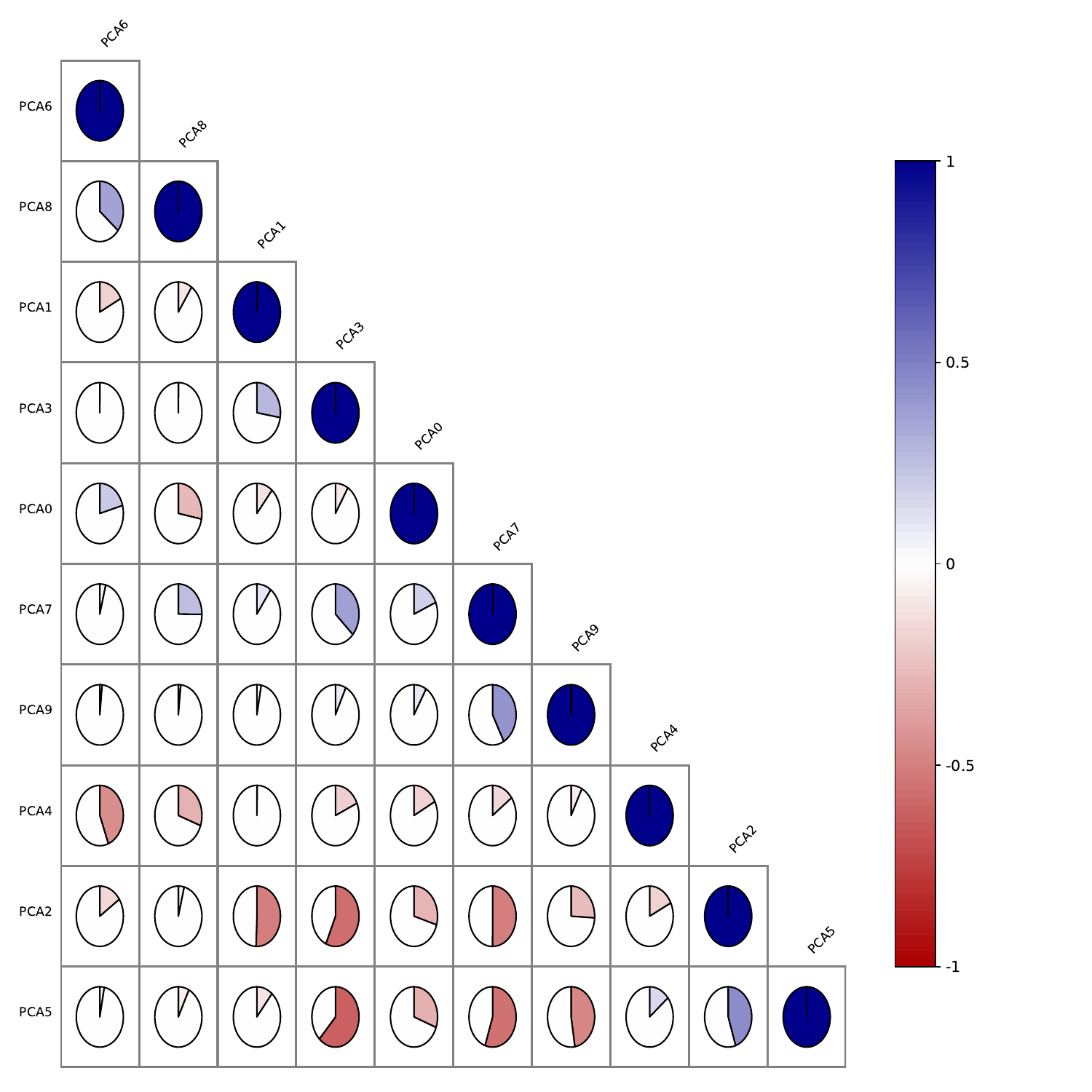}

}

\subcaption{\label{fig-corrleationPlot-withPCA}Correlation between the
principal components after dimensions reduction and model fit}

\end{minipage}%

\caption{\label{fig-raw-pca-dimensionality-reduction}Illustrative
correlation plots to show the strength of association measured by the
Pearson Correlation from the pool of \(100\) experiments. The strength
of relationship is represented by proportion of the gradient of color
ranging between \(-1\) (red) and \(+1\) (blue). The area of circles in
each plot represent the corresponding coefficient of correlation
(\(\rho\)). The left-hand side depicts the formation of clusters among
the ``similar'' variables (Figure~\ref{fig-corrleationPlot-withoutPCA})
created from the raw features which justifies the dimensionality
reduction in the right hand-side
(Figure~\ref{fig-corrleationPlot-withPCA}).}

\end{figure}

Figure~\ref{fig-corrleationPlot-withPCA} is representative subplots of
bivariate relationships between the \(10\) \emph{PCs} from a pool of one
hundred experiments. Each sub-plot shows correlation between defined by
the direction of maximum variance impacted by the linear reduction of
multi-dimensional data into uncorrelated features ordered by the values
of the explained variance of each component. Directions of \emph{PCs}
are not uniquely determined by the component itself, therefore, reading
it from either axis does not alter the meaning of the plot itself.

Correspondingly, the heatmap
(Figure~\ref{fig-corrleationPlot-withoutPCA}) for its intuitive appeal
and ease of summary is organized as sub-plots of pie charts to portray
the strength of association and direction between covariates measured by
the Pearson correlation represented by the color gradient. The deeper
the gradient higher the value. Blue and red gradients respectively
indicate positive (perfect at \(+1\)) and negative (perfect at \(-1\))
correlation. The solid pie across the diagonal represents the perfect
correlation. Figure \ref{fig:corrleationPlot withoutPCA} does not show a
strong outlier that falls outside of a peculiar pattern. A strong
correlation between \emph{idiosyncratic} and \emph{total volatility}
represents the underlying interplay of the components of unsystemic and
systemic risk. It indicates that transactions of insiders tend to
strategize around market volatility (Koumou (2020)). In addition, the
linear direction between the \emph{quick} and \emph{current ratio} means
insiders' short-term liquidity accesses influence anomalous insider
transactions (Cohen, Malloy, and Pomorski (2012)). The negative
correlation between the director and the officer indicates that an
insider can be justifiably a director overseeing the company but may not
engage in the daily operations which is a blurred line and often
unrecorded (Gregory, Matatko, and Tonks (1997)).

\subsection{Results of Classification of Insider Trading
Transactions}\label{sec-analysis-Results-Classification-Transactions}

In this section, we present the performance of the components of the
confusion matrix obtained from the \emph{RF} model fitting. Our results
summarized in Table~\ref{tbl-rfComparativeConfusionMatrixKrishnaResults}
is an average of one hundred experiments each with \(5\)-fold
cross-validation. In our context, the larger number of experiment
smoothed the variation of the performance. We ran experiment ten and
hundred times. As an example standard deviation obtained for accuracy
with ten experiments in case of \(320\) transactions and \(25\) features
respectively of \(10.6\) and \(6.91\) with and without PCA which
declines to \(0.31\) and \(0.45\) respectively for one hundred
experiments. We observed similar results applies to other metrics,
therefore we present results from one hundred experiments.

In Table~\ref{tbl-rfComparativeConfusionMatrixKrishnaResults} we
organize results according to the number of transactions (\(320\) and
\(3984\)), features (\(25\) and \(110\)), and finally sub-divided by
integration with or without \emph{PCA}. The metrics are the
best-reported obtained during a random search in hyperparameter space in
the \(5-\) fold cross-validation settings. For benchmarking, we compare
our results to \emph{DCZ} presented in
Table~\ref{tbl-rfComparativeConfusionMatrixBenchMarkMethod} (Deng et al.
(2021), Deng et al. (2019)) with a focus on the \emph{fifth} column as
proposed by Deng et al. (2021). As an ancillary for the baseline, we
selectively present results from Deng et al. (2019). As shown in
Table~\ref{tbl-rfComparativeConfusionMatrixBenchMarkMethod} (fifth
column), the benchmark method - \emph{PCA-RF} accurately classifies
\(77.88\) percent of transactions. In analogous settings, the lowest
classification accuracy as averaged with one hundred experiments is
\(80.12\) percent (second column of
Table~\ref{tbl-rfComparativeConfusionMatrixKrishnaResults}). In any
other settings, our results beat all benchmarks demonstrated by `ACC'
row of the Table~\ref{tbl-rfComparativeConfusionMatrixKrishnaResults}.
When we increase the count of transactions (fifth-eighth columns,
Table~\ref{tbl-rfComparativeConfusionMatrixKrishnaResults}) the
performance gains the momentum. We credit two explanatory reasons for
the prominence, first, it is influenced by the random selection of
lawful transactions from a larger pool of captured required information,
and second, we unboxed the analysis from time-window perspective. Such
settings has direct potential applicability to the \emph{SEC} to
experiment in the production-level data. Despite keeping the unlawful
transaction unchanged and by randomly sampling the lawful transactions
(\(50\) percent) we demonstrate \emph{RF} beat every benchmark
performance. Overall our results indicate a significant increment in the
accuracy as shown by the \emph{ACC} row of
Table~\ref{tbl-rfComparativeConfusionMatrixKrishnaResults}. Meanwhile,
as a caveat, even though accuracy provides us a basis for basic
inference, literature cautions that taking it as a sole decision-making
gauge may ignite acceptance of the overgeneralized results.

\singlespacing

\begin{longtable}[c]{lrrrrr}

\caption{\label{tbl-rfComparativeConfusionMatrixBenchMarkMethod}Performance
of various metrics according to the benchmark methods on \(320\)
transactions and \(26\) features to identify and detect unlawful insider
trading. Courtsey of Deng et al.~2021 and Deng et al.~2019}

\tabularnewline

\centering

\tabularnewline

\toprule
\multicolumn{4}{c}{ } & \multicolumn{2}{c}{Random Forest} \\
\cmidrule(l{3pt}r{3pt}){5-6}
 & ANN & SVM & Adaboost & No PCA\textsuperscript{a} & With PCA\textsuperscript{*}\\
\midrule
\cellcolor{gray!10}{ACC} & \cellcolor{gray!10}{69.57} & \cellcolor{gray!10}{75.33} & \cellcolor{gray!10}{74.75} & \cellcolor{gray!10}{77.15} & \cellcolor{gray!10}{77.88}\\
FNR & 19.21 & 21.42 & 26.62 & 20.14 & 22.70\\
\cellcolor{gray!10}{FPR} & \cellcolor{gray!10}{34.07} & \cellcolor{gray!10}{27.75} & \cellcolor{gray!10}{24.42} & \cellcolor{gray!10}{25.48} & \cellcolor{gray!10}{21.56}\\
PRE & NA & NA & NA & NA & 78.94\\
\cellcolor{gray!10}{TNR} & \cellcolor{gray!10}{65.93} & \cellcolor{gray!10}{72.75} & \cellcolor{gray!10}{75.58} & \cellcolor{gray!10}{74.52} & \cellcolor{gray!10}{78.44}\\
\addlinespace
TPR & 80.79 & 78.58 & 73.38 & 79.86 & 77.30\\
\bottomrule
\multicolumn{6}{l}{\rule{0pt}{1em}\textsuperscript{a} Deng et al.(2019), No PCA}\\
\multicolumn{6}{l}{\rule{0pt}{1em}\textsuperscript{*} Deng et al.(2021), With PCA}\\

\end{longtable}

\onehalfspacing

Sensitivity or true positive rates (\emph{TPR}) or Recall (\emph{REC})
is a recommended metric for decision-making rather than overall
accuracy. The larger \emph{REC} means that the model reliably detected
the presence of lawful transactions and therefore there exists a lower
number of wrongly classified as unlawful (False Negative) implying
detection was effective with minimum spillage. Among the benchmark
methods \emph{ANN} identifies and classifies \(80.79\) percent lawful
transactions which is marginally higher than \emph{DCZ}'s proposed
method at \(77.30\) percent. On the other hand, our model's ability to
``rule'' lawful as lawful improves evenly on every instance as displayed
by the \emph{TPR} row of
Table~\ref{tbl-rfComparativeConfusionMatrixKrishnaResults}, with the
lowest value of \(84.79\) (second column). \emph{TPR} does not consider
lawful incorrectly identified as unlawful therefore if a data point is
bogus or indeterminate the result becomes useless for detecting or
``ruling in'' true negative. In simpler terms, the \emph{TPR} metric is
useful for ``ruling out'' the possibility of unlawful transactions as it
rarely misidentifies them.

The proposed method reliably detected the presence of lawful
transactions elucidating that wrongful classification of unlawful is low
or vice versa, represented by False Positive Rates (\emph{FPR}) defined
as unlawful transactions incorrectly identified as lawful. Benchmark
methods show that \emph{PCA-RF} FPR of \(21.56\). Our model incorrectly
identifies \(24.55\)
(Table~\ref{tbl-rfComparativeConfusionMatrixKrishnaResults}, second
column). In similar settings of the benchmark methods our proposed
method performs better (columns first through fourth
Table~\ref{tbl-rfComparativeConfusionMatrixKrishnaResults}). We
demonstrate that reducing dimensions is not an effective measure to
detect lawful transactions classified as unlawful. Our proposed model
makes a few mistakes as low as \(1.03\) percent when all features are
considered (Table~\ref{tbl-rfComparativeConfusionMatrixKrishnaResults},
seventh column). Marginally more errors are made by the model when
additional transactions are added which is intuitive. The lower the
value of \emph{FPR} indicates that \emph{RF} can perform better.
Overall, our method beat all \emph{FPR} statistics of benchmark methods
which we showcase in every instance of experiments. It is ``strong
evidence'' of high true positives and low false negatives.

The corresponding metric, specificity is the true unlawful (True
Negative Rates, \emph{TNR}) transactions correctly identified as
unlawful. It excludes the lawful transactions resulting in a higher
number of true unlawful and a low number of unlawful transactions that
are wrongly classified as lawful. That is, the test was successful in
``ruling in'' transactions that were unlawful in every instance when
compared to benchmark methods. Compared to benchmark methods like the
metrics we described above, our performance starts improving as we add
more transactions and/or features. As like other metrics, our metrics
starts with additional transactions and features to surpass all the
benchmark methods. For instance, even when lesser features were
considered than the benchmark method (\(25\), fifth column of
Table~\ref{tbl-rfComparativeConfusionMatrixKrishnaResults}) we show
strength in identifying unlawful transactions as unlawful. Naturally,
our proposed method if integrated with \emph{PCA} has lower performance
among themselves than otherwise but by adding more transactions even
with the PCA integration performance improves by \(18.9\) percent
(difference between \(97.34\) and \(78.44\)). As displayed in
Table~\ref{tbl-rfComparativeConfusionMatrixKrishnaResults} our results
indicate the random selection of positive transactions plays a role in
defining unlawful transactions. Even though metrics suggest a
substantial ability, the result should not be taken alone and be gauged
on potential over-fitting and the existence of noise variables that
remain unaccounted for during data transformation. The result is
inconclusive when lawful transactions are introduced as it cannot ``rule
in'' the false unlawful transactions. However, transactions must be
negative to be identified as negative which will lower the
administrative burden.

In each scenario, our experiment showed strong ability in correctly
identifying unlawful trades. With the selected features and reduced
dimensions, the results are closer to the previous \emph{DCZ} study. It
can be attributed to one of the striking features may be because
inaccessibility of qualitative features for the proposed method.
Moreover, in a notable exception, the \emph{DCZ} study presents results
as a time window. It is interesting as the details of methodology
neither conceive nor justify how temporal properties can be implemented
when results are presented to compare time window length. It can
potentially subject the results to be biased or misclassified. In their
defense of the \emph{DCZ} study, the careful selection of indicators
encompassed explainable features as expounded by economic theories even
with the reduced number of indicators the performance is better.

\singlespacing

\begin{longtable}[t]{lllllllll}

\caption{\label{tbl-rfComparativeConfusionMatrixKrishnaResults}Average
    of the performance metrics of \(100\) experiments each with \(5\)-fold
    cross-validation. Columns represent results with a full set of
    transactions or a subset of the full database matching count of
    transactions and/or features}

\tabularnewline

    \toprule
    \multicolumn{1}{c}{ } & \multicolumn{4}{c}{\makecell[c]{Subset of 320 random selections from \\3984 transactions matching the count \\of the previous study}} & \multicolumn{4}{c}{All available 3984 Transactions} \\
   
    \multicolumn{1}{c}{ } & \multicolumn{2}{c}{25 Features} & \multicolumn{2}{c}{110 Features} & \multicolumn{2}{c}{25 Features} & \multicolumn{2}{c}{110 Features} \\
    
    & No PCA & With PCA & No PCA & With PCA & No PCA & With PCA & No PCA & With PCA\\
    \midrule
    ACC & 82.95 & 80.12 & 90.54 & 83.42 & 97.87 & 97.14 & 99.13 & 98.13\\
    FNR & 14.23 & 15.21 & 7.29 & 13.66 & 1.53 & 1.41 & 0.7 & 1.07\\
    FPR & 19.88 & 24.55 & 11.64 & 19.49 & 2.72 & 4.31 & 1.03 & 2.66\\
    PRE & 81.61 & 77.99 & 89.14 & 81.94 & 97.31 & 95.81 & 98.96 & 97.38\\
    TNR & 80.12 & 75.45 & 88.36 & 80.51 & 97.28 & 95.69 & 98.97 & 97.34\\
    TPR & 85.77 & 84.79 & 92.71 & 86.34 & 98.47 & 98.59 & 99.3 & 98.93\\
    \bottomrule

\end{longtable}

\onehalfspacing

For all features the classification of the true positive (True Legal)
versus the false positive (False Legal) was lower than that of the
selected features. The selected features method has higher probabilities
of detecting true positives than with all features. \emph{AUC} method to
evaluate the model for the best is better in either case however the
value is larger than \(0.5\) (better than random guessing). By comparing
the \emph{AUCPR} across all and selected features it is more evident
that the variation in the values mean of \(13.3\) (\(3984\)) versus
\(98.0\) (\(320\)) provides an inconclusive result. Such a result can be
indicative that selected features are more curated thus it may surpass
the required information thereby inflating the number of
identifications.

The results discussed above are based on the tuned hyperparameters in
\(5\)-folds cross-validation with \(5\) iterations. Theoretically, the
lower values of \(m_{try}\) mean that the performance is better as
opposed to the larger values. The minimum and maximum values of
\(m_{try}\) were respectively \(\textit{0.35}\) and \(\textit{0.95}\)
for \(110\) features with maximum \emph{AUC} of \(0.83\). Increasing the
number of input features increased the \(m_{try}\) but with a marginal
increase in the \emph{AUC}, indicating that features increased the
number of False Positives and the model's performance declined.
According to Probst and Boulesteix (2017), Scornet (2017) that
\(ntrees\) is not a tunable parameter but \emph{RF} expects it to be set
at a high value to obtain better performance, that is, it assists in the
parameter in converging errors. Probst and Boulesteix (2017) empirically
shows that performance gains by growing the number of trees by more than
\(100\). In the study, the maximum and minimum values were \(1030\) and
\(100\) when all \(110\) features were input to the model. When all
features are considered and \(0.98\) (max value) while the max depth is
\(18\). Overall having lower \(sample \ rate\) and \(m_{tries}\) values
leads to the less correlated trees while they provide different
prediction values from each other.

\subsection{Variable Importance}\label{sec-analysis-variable-importance}

\emph{RF} based on a subset of a few strong covariates performs an
implicit variable selection from the underlying latent structure in data
by eliminating irrelevant raw features without diluting classification
accuracy (Genuer, Poggi, and Tuleau-Malot (2010)). For demonstration, in
this section, we choose a sample experiment to illustrate the technique
starting with the original \emph{Gini Scores} proposed by Breiman (2001)
based on the concept of information loss. While Gini impurity scores
provide a strong foundation for explicit feature selection, one might
also be interested in learning the orthogonal splits of the feature
space is also optimal for the classification of the correlated and noisy
features such as finance data. Therefore to observe the distinction we
extend the analysis to include permutation importance.

\subsubsection{Impurity Based Variable
Importance}\label{sec-analysis-variable-importance-impurity-based}

\begin{figure}
     \centering
     \begin{subfigure}[b]{0.49\textwidth}
         \centering
         \includegraphics[width=\textwidth]{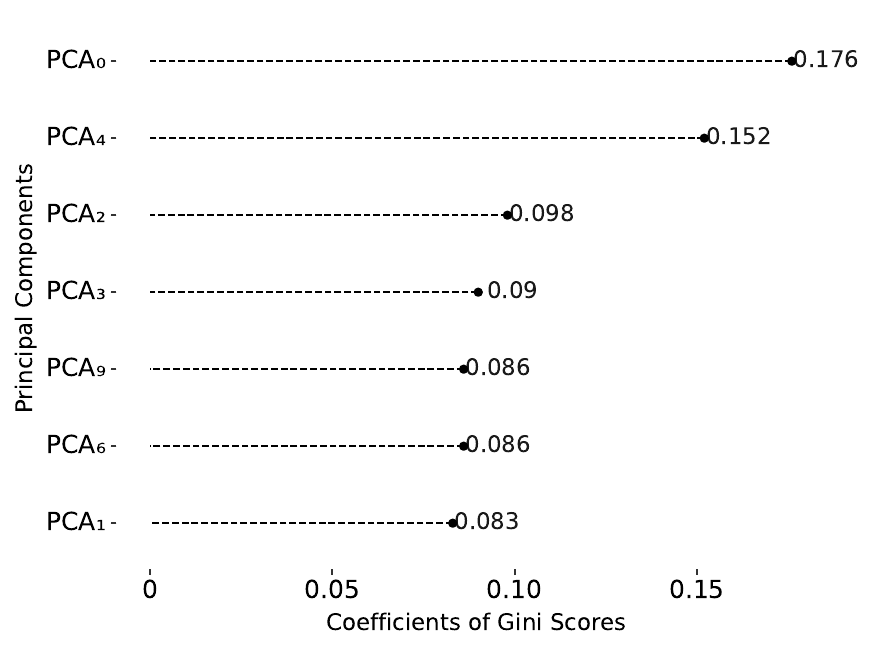}
         \caption{25 Features and 320 Transactions Match Previous Study}
         \label{fig:hyperParam selectedTransactions}
     \end{subfigure}
     \begin{subfigure}[b]{0.49\textwidth}
         \centering
         \includegraphics[width=\textwidth]{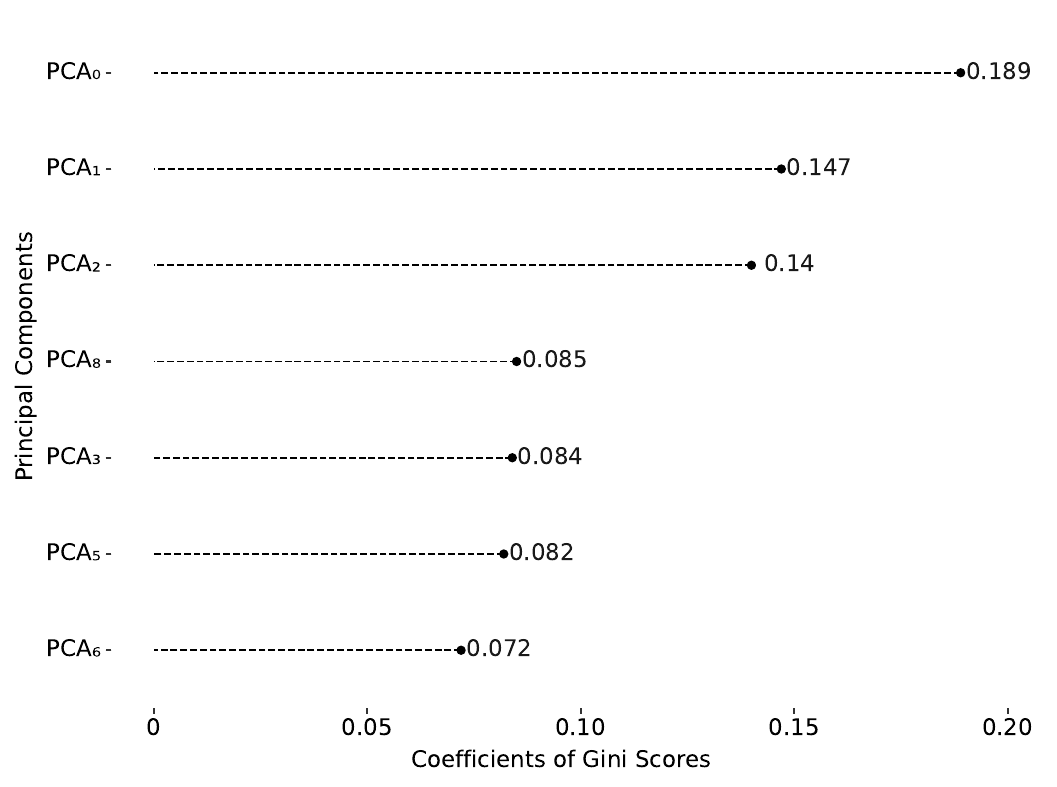}
         \caption{110 Features and 3984 Transactions}
         \label{fig:pcaRank AllFtsAllTrans}
     \end{subfigure} 
     \caption{Ranking of the importance of principal components based on Gini Scores Impurity Scores. The score is the probability of incorrectly classifying a randomly chosen tree if the tree were labeled according to the class of distribution which is equivalent to log loss. It is given by $G=\sum_{i=1}^C p(i) \times (1-p(i)) \text{where C is number of classes for instance i, p(i) is the probability}$ }  
\end{figure}

We randomly choose a sample experiment to compare the feature ranking
based on the \emph{MDI} to ranking of the Deng et al. (2021). In the two
parts visualizations - \ref{fig:hyperParam selectedTransactions} ranks
the principal components for the \(25\) followed by
\ref{fig:pcaRank AllFtsAllTrans} for \(110\) features. The horizontal
and vertical axis of \ref{fig:hyperParam selectedTransactions} and
\ref{fig:pcaRank AllFtsAllTrans} respectively represent coefficients of
\emph{Gini Scores} and the principal components in the descending order.
In \ref{fig:hyperParam selectedTransactions} \(\text{PCA}_0\) has the
highest and \(\text{PCA}_1\) lowest ranks with other components in
between in the order of
\(\text{PCA}_0 > \text{PCA}_4 > \text{PCA}_2 > \ldots \text{PCA}_1\) for
\(25\) features. The coefficient of \emph{Gini Score} refers to how
often a particular feature was selected for a split. Hence, the highest
coefficient score represents the larger contribution to the split
ranking it higher. According to
Table~\ref{tbl-pcacompartivdddeFeatureImportranceRF} (a) for
\(\text{PCA}_0\), the largest contribution was made by returns and
acquisition-disposition which is intuitive given the asset purchase and
sell activities by insiders. It is well-known in finance literature
returns explain the largest variance as investors' choices in the face
of uncertainty about the future (Foresi and Peracchi (1995)).
Risk-related features related to the fair market pricing value and
attributed to playing a sensitive role in determining insider's
transactions and ultimate behavior.

Likewise, \(\text{PCA}_0\)- the highest ranking principal component in
Table~\ref{tbl-pcacompartivdddeFeatureImportranceRF} (b) for all
features is influenced by its returns, alpha, and beta - all covariates
related to the risk-taking behavior. In \ref{fig:pcaRank AllFtsAllTrans}
\(\text{PCA}_0\) and \(\text{PCA}_6\) respectively represent the highest
and the lowest values of coefficients in descending order of
(\(\text{PCA}_0 > \text{PCA}_1 > \ldots \text{PCA}_6\) ). Evidently,
return is related to the risk-performance of the asset which supports
the idea that insiders exert more influence in determining asset prices
that ultimately can influence unlawful trading because of short-term
motives. These features are highly cardinal and correlated and may not
be as influential as they seem to be. In contrast, it is intuitive to
think that the influence of executive insiders may impact companies more
than quarterly returns or market risk components reflected in unlawful
transactions.

\begin{table}

\caption{\label{tbl-pcacompartivdddeFeatureImportranceRF}An illustrative
example of the contribution of individual covariates in the order of
their importance (left to the right) to determine the principal
components. Table (a) and (b) represent coefficients calculated for
selected \(25\) and \(110\) features and for \(320\) transactions. PCs
in columns are organized according to their order of ranking based on
Gini Scores (highest to lowest, left to the right).}

\begin{minipage}{0.50\linewidth}

\centering
\resizebox{\ifdim\width>\linewidth\linewidth\else\width\fi}{!}{
\begin{tabular}{lrrr}
\toprule
Features & $PCA_0$ & $PCA_4$ & $PCA_2$\\
\midrule
Return & 0.32 & 0.02 & -0.11\\
Acq. Disp. & 0.12 & -0.39 & -0.32\\
Sprd. Rtn. & -0.34 & -0.16 & -0.16\\
Market Beta & -0.31 & 0.15 & -0.17\\
SMB Beta & -0.02 & 0.40 & 0.16\\
HML Beta & -0.37 & 0.07 & -0.34\\
Idiosyncratic Volatility & -0.39 & 0.06 & -0.34\\
Total Volatility & -0.01 & -0.26 & 0.09\\
Prc. Op. Earnings(Basic) & -0.16 & -0.49 & 0.04\\
Price Book & -0.29 & 0.02 & 0.28\\
\bottomrule
\end{tabular}}

\end{minipage}%
\begin{minipage}{0.50\linewidth}

\centering
\resizebox{\ifdim\width>\linewidth\linewidth\else\width\fi}{!}{
\begin{tabular}{lrrr}
\toprule
Features & $PCA_0$ & $PCA_1$ & $PCA_2$\\
\midrule
Return & 0.14 & -0.14 & 0.03\\
Alpha & -0.02 & 0.04 & -0.02\\
Market Beta & -0.10 & 0.06 & 0.03\\
SMB Beta & -0.06 & 0.13 & -0.05\\
HML Beta & -0.02 & -0.02 & -0.04\\
Idiosyncratic Volatility & -0.05 & 0.10 & -0.05\\
Total Volatility & -0.06 & 0.11 & -0.05\\
R Squared & -0.17 & 0.03 & 0.07\\
Excess Returns & 0.01 & -0.06 & 0.02\\
IsDirector & -0.01 & 0.00 & 0.00\\
\bottomrule
\end{tabular}}

\end{minipage}%

\end{table}%

\subsubsection{Permutation Based Variable
Importance}\label{sec-analysis-variable-importance-permutation-importance}

Even though impurity-based feature importance ranks the principal
components based on the correlation as shown in
Table~\ref{tbl-pcacompartivdddeFeatureImportranceRF}, it is hard to
infer the rank of the features directly based on these scores as they
vary from one to other components. Besides, the \emph{MDI} based methods
are biased towards high cardinal features and its ranking is based on
training features therefore does not typically generalize to the test
set. Additionally, our dataset belongs to finance and therefore contains
highly correlated covariates, and one variable may contain information
to classify the response variable (Meinshausen (2008)). When one of the
features is permuted, the model still has access to both features
through correlation which results in lower relative importance for both
features, even though both of them may be significantly important. To
overcome these shortcomings, in this section, we rank features based on
the permutation importance, a model-agnostic method that can applied to
any fitted estimator by calculating the value of the contribution of
each \(m^{th}\) variable in the out-of-bag examples keeping all other
predictor variable fixed and passed on to other corresponding trees
(Nembrini, König, and Wright (2018)). We construct hierarchical clusters
based on Spearman rank correlation\footnote{Spearman rank correlation is
  an appropriate method because the correlation coefficients will be
  higher if two variables have similar (or identical) ranks, and lower
  if they have lower ranks.} of the highly correlated covariates, track
one representative feature from each cluster and finally permute between
them. The linkage between the clusters is established based on the
Ward's minimum variance. The distance matrix is the derivative of the
correlation matrix.

\begin{figure}
     \centering
     \begin{subfigure}[b]{0.49\textwidth}
         \centering
         \includegraphics[width=\textwidth]{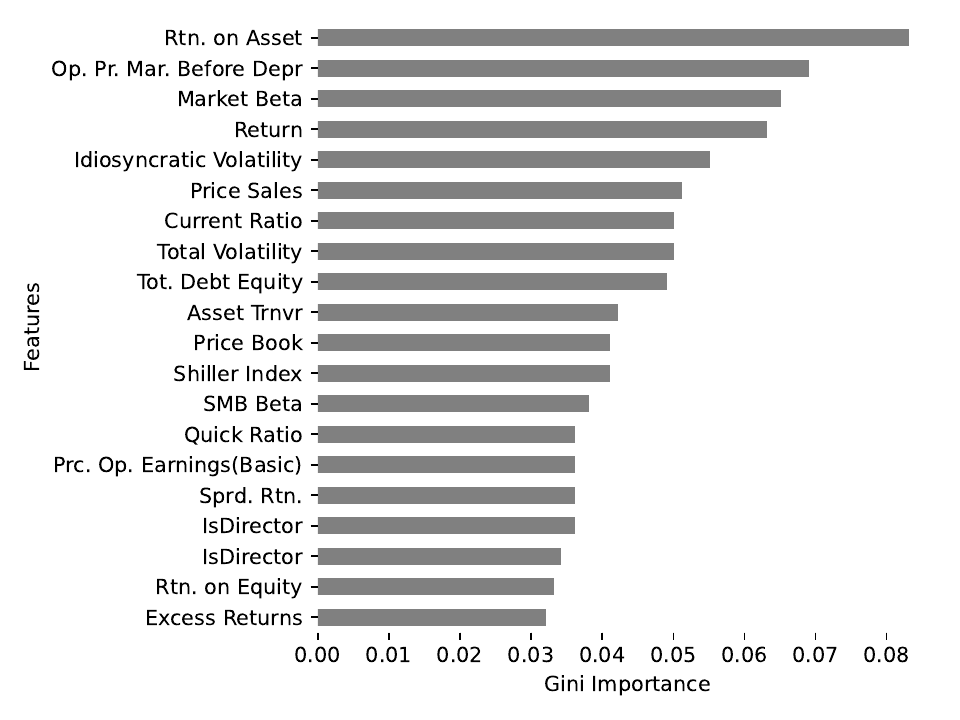}
         \caption{Gini Scores}
         \label{fig:varImp giniImpPermImp}
     \end{subfigure}
     \hfill
     \begin{subfigure}[b]{0.49\textwidth}
         \centering
         \includegraphics[width=\textwidth]{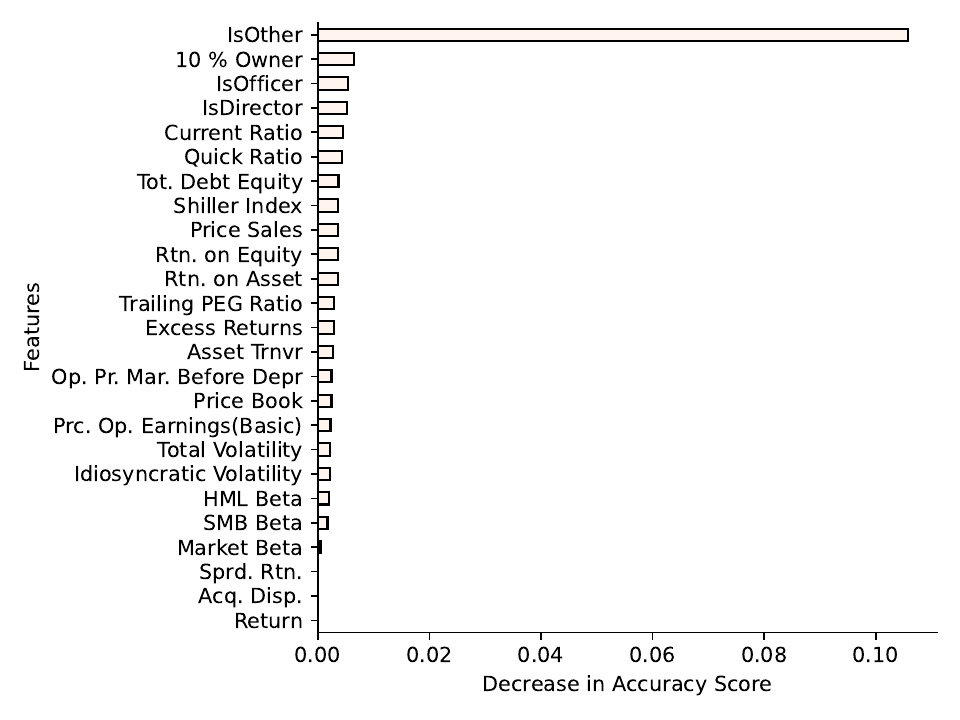}
         \caption{Permutation Values}
         \label{fig:varImp permImpTest}
     \end{subfigure} 
     \caption{Tranining Data: Figure \ref{fig:varImp giniImpPermImp} shows the ranking of features based on the Gini scores based method described by \cite{breiman2001random}. Figure \ref{fig:varImp permImpTest} shows the ranking based on permutation importance. Both plots are are randomly choosen from a set of one hundred exepriments with $3984$ transactions and $110$ features.}
\end{figure}

In Figure \ref{fig:varImp giniImpPermImp} we rank features based on the
Gini Impurity Scores. In Figure \ref{fig:varImp giniImpPermImp} top
variables such as return on asset, operating profit margin before
depreciation, and market \(\beta\) tend to have multiple distinct values
which are demonstrative that \emph{MDI} is biased towards them leading
to a higher rank. Notably, RoA and Profit Margins are highly correlated
as well as market \(\beta\) and returns which complicates decisions
based on visual inspection. In contrast by performing permutation
importance in the same training dataset (Figure
\ref{fig:varImp permImpTest}), we observe that low cardinal categorical
features such as ``IsOther'', ``Ten Percent Owner'' and ``IsOfficer''
are the most important features. Even before removing correlation if we
permute at the most there is a decline of \(0.10\) in accuracy which
suggests none of the features are important. This is in contradiction to
our high very test accuracy averaging at \(99.13\) (with PCA), \(98.13\)
(without PCA) in
Table~\ref{tbl-rfComparativeConfusionMatrixKrishnaResults}. Therefore to
verify the change in accuracy is caused by the cardinality and highly
correlated covariates, we perform hierarchical clustering and rank the
features based on the Spearman Rank Correlation and re-run permutation
importance in the held-out set.

\begin{figure}
     \centering
        \includegraphics[width=0.90\textwidth]{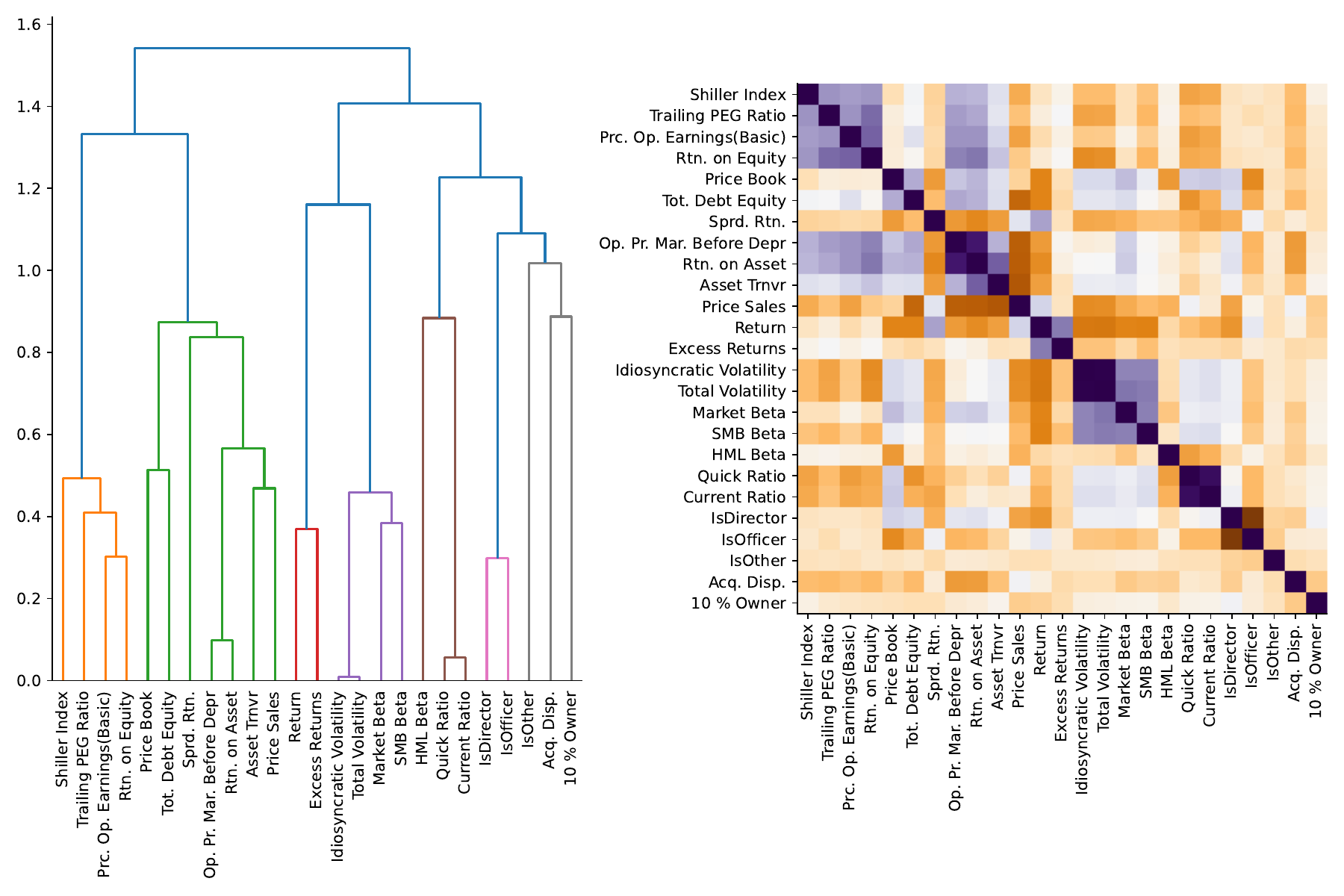}
         \caption{Hierarchical clustering of the features based on the Spearman rank-order correlations and the heatmap of the correlated features. The left-hand shows the dendogram - a graphical representation between the sequentially merged pairs of similar covariates forming neighbors to each other. The right-hand side is a correlation plot based on Spearman rank-order correlations of few selected features shown for illustrative purposes only}
         \label{fig:varImp dedogramCluster}     
\end{figure}

In the right-hand side of Figure \ref{fig:varImp dedogramCluster} we
display a correlation matrix heatmap of the selected features with a
diagonal showing the perfect correlation. The left side of Figure
\ref{fig:varImp dedogramCluster} shows the dendrogram reflecting the
relationship of similarity among the group of variables connected by
\textit{clades}. For instance, Price Earnings (Basic) and Return on
Equity(RoE) form a clade which joined with the Trailing PEG Ratio forms
a different clade for the leftmost clade in Figure
\ref{fig:varImp dedogramCluster}. Analogously, the rightmost cluster
acquisition and Disposition and Ten Percent Ownership form a clade.
Therefore the arrangement of each of these clades shows which features
(leaves) are similar to each other distinguished by the height - the
higher the height, the greater the difference.

\begin{figure}
     \centering
     \begin{subfigure}[b]{0.49\textwidth}
         \centering
         \includegraphics[width=\textwidth]{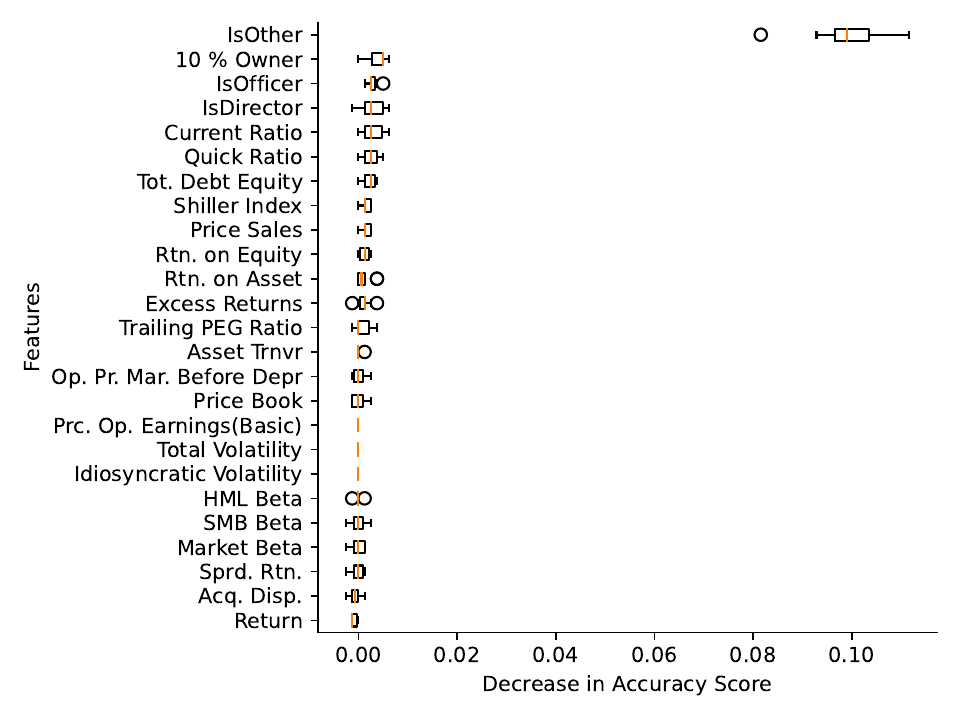}
         \caption{Ranks before removal of multicollinearity}
         \label{fig:varImp permImpMulTest}
     \end{subfigure}
     \hfill
     \begin{subfigure}[b]{0.49\textwidth}
         \centering
         \includegraphics[width=\textwidth]{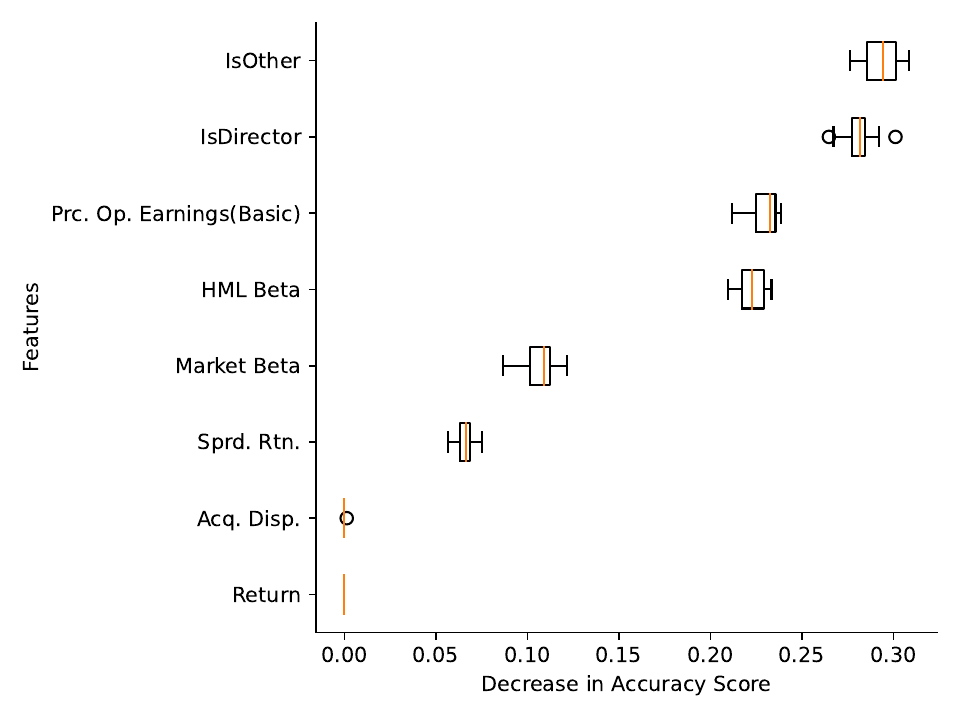}
         \caption{Ranks after removal of multicollinearity}
         \label{fig:varImp permImpMulRem}
     \end{subfigure}    
     \caption{Test Data: Ranking of the covariates based on descending order of the permutation values. The Figure \ref{fig:varImp permImpMulTest} is the ranking before the hierarchial clustering of covariates based on the Spearman rank-order correlation as shown in Figure \ref{fig:varImp dedogramCluster}. The ranking shown in \ref{fig:varImp permImpMulRem} is based on permutation values obtained after removing collinearity} 
\end{figure}

By removing correlation we can observe the accuracy can drop by as much
as by \(0.3\) (Figure \ref{fig:varImp permImpMulRem}) versus
uncorrelated (\(0.10\) as shown in Figure
\ref{fig:varImp permImpMulTest}). The distinction of ownership and
institutional covariates such as ``IsOther'' did not change but its
prominence is more visible after decorrelation. We agree with Strobl et
al. (2008) stating that permutation importance does not break the
relationships as with correlated predictor variables is prominent,
especially in the higher dimensional problems. In general, the features
of financial statements play an important role in direct activities in
the stock market even so. Governance of the company and activities in
purchase and sales play important roles. It can be inferred that even
though variables like \emph{acquisition-disposition} are at the bottom
of the ranking they exert influence. Figures
\ref{fig:varImp permImpMulTest} and \ref{fig:varImp permImpMulRem} show
that Ownership cannot be overshadowed by the activities and/or risks
taken in the market. The market \(\beta\), \emph{value premium} (High
Minus Low) - spread between the companies with high book-to-market
ratios (value companies) outperform the companies with low
book-to-market (value) also play important roles, which given the
institutional influence of the executives play important roles in
deciding an activity lawful and/or unlawful (e.g.~dividend policy,
Campbell and Shiller (1988), Grinblatt, Masulis, and Titman (1984),
Henry, Nguyen, and Pham (2017)).

\section{Discussion}\label{sec-analysis-discussions}

We proposed two variations of \emph{RF} to detect unlawful insider
trading and elicit relevant features based on interpretable statistical
scores from detailed trades and financial data. With balanced accuracy
and sensitivity in mind, we evaluated the performance of each model in
hyperparameter-tuned settings. In addition, we replicated steps in the
equivalent method proposed by \emph{DCZ} for meaningful comparisons. We
extended the \emph{DCZ} method by semi-manually labeling data, randomly
selecting transactions from multiple industries, removing undefined
qualitative features, and fed into an automated pipeline to train and
test the fitted model. Our results demonstrate that the model
successfully uncovered latent structures in the data, gained better
performance than \emph{DCZ} for each method and boosted the
interpretability. We can confirm that governance, financial and trading
features may be relevant and only a few selected variables may naturally
contribute to uncovering fraudulent behaviors. Our assessment avoided
optimistical biased performance with the introduction of additional
features indicating that the model can concretely learn and distinguish
relevant unlawful transactions. This was achieved by implementing a
range of hyperparameters in \emph{5-fold} cross-validation by minimizing
generalization error. Therefore, we extend the results obtained by the
\emph{DCZ} study with the following highlights:

\begin{itemize}
\item
  \emph{Low Fall Out Rate:} In detecting unlawful transactions, it is
  very important that the proposed method minimizes mistakes to identify
  unlawful transactions. As an analogous to a situation when a person
  guilty of a crime is acquitted, comparing \emph{FPR} with benchmark
  methods in
  Table~\ref{tbl-rfComparativeConfusionMatrixBenchMarkMethod}, the
  proposed method commensurate proportionate or better performance. The
  model successfully minimizes false alarms by not wrongfully
  classifying unlawful as lawful. Because \emph{FPR} metrics are not
  controlled by researchers it reduces arbitrary judgment and solves the
  limitations of traditional econometric studies associated with \emph{a
  priori} settings of arbitrary significance levels. Such finding is
  important to the SEC to confide \emph{ML} methods to relieve labor
  intense activities to redirect to enhance rule-making to tracking of
  the uncaptured unlawful insider trading transactions.
\item
  \emph{Low Miss Rates:} By having lower \emph{miss rates}, the model
  makes very few mistakes to classify lawful as unlawful transactions.
  Analogous to a situation when a non-criminal is incarcerated, if a
  lawful transaction is investigated as unlawful, the burden and stakes
  are high. Our proposed method reports it misses only a few instances
  and wrongfully classifies lawful as unlawful. However, our method is
  substantially better to flag false negatives.
\item
  \emph{High Sensitivity:} The proposed method reliably detected the
  presence of lawful transactions which translates into that wrongful
  classification of unlawful is low or vice versa. This is ``strong
  evidence'' of higher true positives and low false negatives. As a
  starting point, the enforcement agencies can build robust systems and
  refine methods to further improve the performances thus they can focus
  on detailed ramifications of unlawful activities.
\item
  \emph{High Specificity:} The model correctly identified unlawful as
  unlawful which excludes the lawful transactions resulting higher
  number of true unlawful (true negatives) and a low number of unlawful
  transactions that are wrongly classified as lawful.
\item
  \emph{High Accuracy:} In each scenario, the effectiveness as measured
  by the accuracy of classification, the proposed method correctly
  identifies lawful and unlawful transactions better than each of the
  benchmark methods. We demonstrated that a larger pool of data and
  unboxing time-window archives better accuracy which is pragmatic for
  the \emph{SEC} to experiment in the production level data. Despite
  keeping the unlawful transaction unchanged and by randomly sampling
  the lawful transactions (\(50\) percent) we demonstrate \emph{RF} beat
  every benchmark performance. In comparison, the method in each
  instance shows meaningful improvement to Deng et al. (2021).
\item
  In addition to the algorithm prescribed \emph{MDI} based feature
  rankings, we extended it to a model agnostic permutation based
  variable importance. The former includes rankings only on the training
  while the latter has ability to extend to the test dataset. That
  difference allowed us to investigate influence of correlations. By
  doing so we observed that categorical features related to corporate
  governance gain more prominence which were less influential in
  \emph{MDI} based ranking (Figure \ref{fig:varImp permImpMulTest}
  before versus Figure \ref{fig:varImp permImpMulRem} after removing
  correlation). That is, when features are correlated ranking them
  introduces of the possibility of one variable may contain information
  that classifies the response variable therefore the ranking is
  influenced and its prominence is diluted (Avanzi et al. (2023)).
\end{itemize}

\section{Conclusions and future
directions}\label{sec-conclusions-future}

The research analyzed variations of Random Forest in semi-manually
labeled insider trading transactions that reflect ownership, corporate
governance, financial performance, and returns in the security market.
Our research is integration of multiple techniques from the natural
language processing to extract proper names and merge individual
transactions from Form \(4\) based on the Levensthein distance,
independent data labeling, identifying finance and trade covariates
belonging to individual insider transactions to create a persistent
database. Between \(2003\) and \(2022\), we balanced lawful and unlawful
by randomly choosing \(320\) to match benchmark research and continue to
extend the number of transactions (e.g.~\(3984\)) and features
(\(110\)). To control variability we repeated our experiments one
hundred times. While unlawful transactions remains fixed at a maximum of
\(1992\), we randomly choose lawful transactions from a pool of \(9.6\)
million in every iteration. Performance of each instance of model
configuration is sampled in \(5\)-fold cross-validation with
\(5\)-iterations initialized, controlled and optimized by random space
search of the number of parameters to provide evidence that it has
sustained the robustness check and extended on the previous study to
accurately identify and discover unlawful transactions with lower fall
out and miss rates.

Certain features dominate unlawful behavior that, \emph{ex officio},
corporate insiders predominantly are at avail to manipulate, which in
all likelihood encourages committing fraud. Towards inferring that, we
recognized several challenges and recommend for future improvements.
First, decision trees are effective classification methods with the high
accuracy, an extension of these methods to explain the causal effects
are of interest to the regulators, policy makers and researchers (Athey
(2019)). Second, the study is based on the uniform random sampling for
parameter tuning, a simple but potentially wasteful method as the
features space grows. Providing results from alternatives such as
Bayesian Optimization, Grid Search, Evolutionary and so on that could
scale to the complex yet sample-efficient methods is another future
direction (Probst, Bischl, and Boulesteix (2018)). Third, while the
current research extends features from \(25\) to \(110\), in the future,
an extension to include \(447\) anomaly-related features from Kewei. Hou
(2017) may provide more insightful inferences. Fourth, even though
categorical features are handled via one-hot encoding as popular
literature suggests issues arise as the cardinality expands - the
independence (that is, orthogonality) assumption of one-hot may no
longer hold as the feature space becomes crowded and correlation between
the features is inevitable. Fifth, our results are based on a balanced
dataset which may not be very pragmatic even though we tried to perform
several experiments with changing transactions, it would become more
robust if the results from the unbalanced dataset match. Overall, our
results demonstrate that methods are efficient techniques that fuse data
mining and modeling to extract predictive features in a unified
framework to derive robust results enabling cheaper, faster and creative
ways of \emph{data labeling}, automating routine tasks by uniquely
positioning to capture \emph{short-run noise} which leads to
\emph{trustworthy} predictive power\footnote{Prediction is defined as
  the ability to take known information to generate new information
  (Agrawal, Gans, and Goldfarb (2019))} for \emph{understandable
inferences} (Khademian (2022), Igami (2018), Christensen, Hail, and Leuz
(2016)).

\newpage{}

\section*{Appendix}\label{appendix}
\addcontentsline{toc}{section}{Appendix}

\begin{table}

\caption{\label{tbl-Variables}Compiled list of the selected financial
indicators found in financial statements - balance sheet, income and
cash flow. The quantitative covariates are grouped to asess company's
profitability, valuation, liquidity, captialization, financial soundess,
growth, leverage, risk and returns, and governance. (*represent variable
that matches Deng et. al.~2021 and Deng et. al.~2019)}

\begin{minipage}{\linewidth}

\begin{longtable*}[l]{>{\raggedright\arraybackslash}p{4cm}>{\raggedright\arraybackslash}p{13cm}}
\toprule
Group & Variables\\
\midrule
\endfirsthead
\multicolumn{2}{@{}l}{\textit{(continued)}}\\
\toprule
Group & Variables\\
\midrule
\endhead

\endfoot
\bottomrule
\endlastfoot
Activity/Efficiency Ratios & Asset Turnover*, Inventory Turnover, Payables Turnover, Receivables/Current Assets\\
Annual Valuation Ratios & Shiller's P/E, Dividend Yield, Dividend Payout Ratio, Enterprise Value Multiple, Price-to-Cash Flow, Price-to-Earnings, excl. EI (diluted)*, Price-to-Earnings, incl. EI (diluted)*, Forward P/E to 1-year Growth (PEG) ratio*, Forward P/E to Long-term Growth (PEG) ratio*, Trailing PEG Ratio, Price-to-Sales Ratio*\\
Capitalization Ratios & Capitalization Ratio, Long-term Debt/Invested Capital, Common Equity/Invested Capital, Total Debt/Invested Capital\\
Financial Soundness Ratios & Cash Flow to Debt, Cash balance to Total Liabilities, Current Liabilities as Percentage of Total Liabilities, Total Debt as percentage of Total Assets, Gross debt to EBITDA, Long-term Debt/Book Equit, Free Cash Flow/Operating Cash Flow, Interest as Percentage of Average Long-term Debt, Interest as Percentage of Average TotalDebt, Inventory/Current Assets, Long-term Debt/Total Liabilities, Total Liabilities/Total Tangible Assets, Operating Cash Flow to Current Liabilities, Profit before D\&A to current liabilities, Receivables Turnover, Short-Term Debt/Total Debt\\
Liquidity Ratios & Cash Conversion Cycle, Cash Ratio, Current Ratio*, Quick Ratio (Acid Test)*, Quoted Spread\\
\addlinespace
Miscellaneous Ratios & Accruals/Average Assets, Advertising as Percent of Sales, Market Capitalization , Price-to-Book Ratio*, Research and Development as percent of Sales, Sales per Dollar Total Stockholders’ equity, Sales per Dollar Invested Capital, Sales per Dollar Working Capital, Labor Expenses/Sales\\
Ownership /Governance & Acquisition Disposition, Derivatives Held, Adjusted Derivatives Held, IsDirector, IsOfficer, IsOther, Ten Percent Ownership\\
Profitability Ratios and Rates of Return & After Tax Return on Average Common Equity, After Tax Return on Total Stock Holder's Equity, After Tax Return on Invested Capital, Alpha (Excess Return), Cash Flow Margin, Effective Tax Rate, Trailing PEG Ratio, Gross Profit Margin*, Gross Profit/Total Assets, Net Profit Margin, Operating Profit Margin After Depreciation*, Operating Profit Margin Before Depreciation*, Pre-tax Return on Total Earning Assets, Pre-tax return on Net Operating Assets, Pretax Profit Margin, Return on Assets*, Return on Capital Employed, Return on Equity*\\
Risk & Ask, Ask or High Price, Beta (High Minus Low)*, Market Beta*, Small-minus-big Size factor*, Bid, Bid Ask Spread, Bid or Low Price, Effective Spread, Excess Return from Risk Model*, Idiosyncratic volatility from the q-factor model, Kyle Lambda, Number of Derivatives, Number of Derivatives after Trade, Number of Trades, Price Impact, Market R-Squared, Realized Spread, Return, Returns without Dividend, Underlying Market Equity Volume, Underlying Shares Adjust, Outstanding Shares, Underlying Market Price, Underlying Market Price Adjust, Spread of Return, Total Volatility*, Volume, Exercise Price, Exercise Price Adjust\\
Shareholder's Equity, Invested Capital and Operating Cash Flow & Book-to-Market\\
\addlinespace
Solvency Ratios & Debt-to-equity Ratio, Debt-to-assets, Debt-to-Capital, After Tax Interest Coverage, Interest Coverage Ratio\\
Valuation Ratios & Price-to-Operating EPS, excl. EI (basic), Price-to-Operating EPS, excl. EI (diluted)\\*
\end{longtable*}

\end{minipage}%

\end{table}%

\newpage{} \# References \{.unnumbered\}

\phantomsection\label{refs}

\end{document}